\documentclass[11pt]{article}

\usepackage[T1]{fontenc}
\usepackage[utf8]{inputenc}
\usepackage{lmodern}
\usepackage{microtype}

\usepackage{cite}
\usepackage{graphicx}
\usepackage{url}
\usepackage{amsmath,amssymb,amsfonts}
\usepackage{bm}
\usepackage{textcomp}
\usepackage{xcolor}
\usepackage{calc}
\usepackage{mathrsfs}
\usepackage{booktabs}
\usepackage{xparse}
\usepackage{xspace}
\usepackage{tikz}
\usetikzlibrary{arrows.meta,calc,positioning,fpu}

\DeclareUnicodeCharacter{00A0}{~}

\newif\ifrevisions
\revisionstrue

\ifrevisions

\else

\fi

\usepackage[hidelinks]{hyperref}
\hypersetup{
	pdftitle={On the Unification of Deterministic and Stochastic Electromagnetic Information Theory via Symplectic Geometry},
	pdfauthor={Marco Donald Migliore},
	pdfsubject={Electromagnetic Information Theory},
	pdfkeywords={Electromagnetic Information Theory, Degrees of Freedom, MIMO, Antennas, Electromagnetic propagation, Communication systems, Massive MIMO, Stochastic systems, Symplectic geometry}
}

\newcommand{\NDF}{\texorpdfstring{\text{NDF}\xspace}{NDF}}
\newcommand{\NDFeps}{\texorpdfstring{\text{NDF}$_{\varepsilon}$\xspace}{NDF epsilon}}

\newcommand{\bE}{\mathbf{E}}
\newcommand{\bJ}{\mathbf{J}}

\newcommand{\bu}{\mathbf{u}}
\newcommand{\br}{\mathbf{r}}

\newcommand{\E}{\mathbb{E}}

\DeclareMathOperator{\sinc}{sinc}

\newcommand{\bpsi}{\boldsymbol{\psi}}
\newcommand{\dG}{\bar{\bar{\mathbf{G}}}}

\newcommand{\dI}{\bar{\bar{\mathbf{I}}}}

\newcommand{\WDFP}{\mathcal{W}_p}

\newcommand{\Vol}{\text{Vol}}

\newcommand{\NDFepsv}{\texorpdfstring{$\text{NDF}_{\varepsilon}$}{NDF epsilon}}
\newcommand{\SIFs}{\texorpdfstring{SIFs\xspace}{SIFs}}
\newcommand{\SIF}{\texorpdfstring{SIF\xspace}{SIF}}

\begin{document}
	
	\title{On the Unification of Deterministic and Stochastic Electromagnetic Information Theory via Symplectic Geometry}
	
	\author{
		M.~D.~Migliore%
		\thanks{M. D. Migliore is with the DIEI, Department of Electrical and Information Engineering ``Maurizio Scarano'', University of Cassino and Southern Lazio, Via Di Biasio 43, I-03043 Cassino, Italy. E-mail: mdmiglio@unicas.it.}%
		\thanks{He is also with the European University of Technology (EUT+), CNIT, Consorzio Interuniversitario per le Telecomunicazioni, and Eledia Center.}
	}
	
	\date{}
	
	\maketitle
	
	\begin{abstract}
		This paper unifies deterministic and stochastic Electromagnetic
		Information Theory (EIT) through symplectic geometry. For spatially
		incoherent sources, both formulations yield identical eigenvalues and
		spatial Number of Degrees of Freedom (NDF). In the asymptotic regime
		and in the absence of losses, this equivalence is shown to be a
		structural necessity: the radiometric \'etendue, the Hamiltonian
		phase-space volume, and the NDF are the same symplectic invariant of
		the source--observer configuration. Liouville's theorem guarantees
		conservation of the NDF under lossless propagation, while Gromov's
		Non-Squeezing Theorem establishes a minimum phase-space cell, setting
		a fundamental geometric bound on resolving power. The physical
		manifestation of this symplectic structure is the formation of
		\textit{Spatial Information Flows} (SIFs), defined operationally as
		the spatial loci along which the spatial coherence, equivalently the
		mutual information, decays at the minimum possible rate. Spatial
		information in electromagnetic fields is therefore governed by the
		geometry of the source--observer configuration, providing the
		foundation for a geometric theory of electromagnetic information.
	\end{abstract}
	
	\begin{center}
	\end{center}
	

\section{Introduction}

In the early development of information theory, two distinct and
competing paradigms emerged. One was grounded in a fundamentally
physical perspective, while the other was rooted in a rigorous
mathematical formalism. The foremost representative of the former was
Dennis Gabor, who laid the foundations of what might be termed a
``physical theory of communication''~\cite{gabor1953communication}.
In contrast, Claude Shannon pioneered the ``mathematical theory of
communication''~\cite{Sha:48}. Shannon's stochastic approach quickly
became the dominant framework in modern information science, thanks to
its ability to model the probabilistic nature of information.

Several decades later, the development of \emph{Electromagnetic
	Information Theory} (EIT) revived a comparable dichotomy. On one
hand, a deterministic approach emerged, based on classical
electromagnetism, identifying the fundamental spatial number of
degrees of freedom (\NDF) through the singular value decomposition
of the radiation operator~\cite{bucci1989degrees, bucci1987spatial,
	migliore2006role, miller1998spatial, poon2005degrees,
	migliore2008electromagnetics, gruber2008new, janaswamy2011degrees,
	franceschetti2015information, franceschetti2017wave,
	gustafsson2025degrees, gradoni2015phase}. On the other hand, a stochastic
framework accounted for the probabilistic nature of
information-carrying electromagnetic fields, characterising spatial
information through the coherent modes of the cross-spectral
density~\cite{mandel1995optical, gradoni2018stochastic,
	pizzo2022spatial, wan2023mutual}.

Despite their different mathematical foundations, both approaches
converge, in the asymptotic regime, to the same answer: the \NDF\
reduces to a purely geometric quantity determined solely by the
source--observer configuration, independently of the specific source
model. This convergence is not limited to the det\-er\-min\-is\-tic
and stochastic EIT formulations: functional approximation
theory~\cite{franceschetti2015landau, slepian1962prolateIII} and
rate-distortion theory arrive at the same geometric limit through
entirely different mathematical routes. The persistent effectiveness
of deterministic methods in EIT~---despite neglecting the stochastic
nature of the problem and relying on tools developed for field
representation well before the emergence of EIT~---is a further
manifestation of the same phenomenon.

By showing that stochastic and deterministic models converge to a
common geometric skeleton, this work identifies symplectic geometry
as the unifying structure underlying the observed equivalence.

The paper develops this argument at three levels. At the
\textit{functional level}
(Sections~\ref{sec:non_stocastic}--\ref{app:functional_models}),
we show that deterministic and stochastic EIT formulations yield
identical eigenvalues, basis functions, and \NDF\ for spatially
incoherent homogeneous sources, and that in the asymptotic regime
both collapse onto the same geometric skeleton in Hilbert
space~---the hyperellipsoid defined by the singular values of the
radiation operator. The same collapse is shown to occur in
functional approximation theory, indicating that the convergence
is not specific to any particular EIT formulation but reflects a
deeper structural property. At the \textit{geometric level}
(Section~\ref{sec:EIT_modelB}), this skeleton is identified with
the symplectic geometry of the source--observer phase space.
This framework formalizes the equivalence between quantities
traditionally treated in distinct literatures: the \'etendue of
radiometry, the phase-space volume of Hamiltonian mechanics \cite{gradoni2015phase}, the shadow-area approach \cite{gustafsson2025degrees}, and
the \NDF\ of information theory. By unifying these perspectives,
the \NDF\ is established as a symplectic invariant of the radiation
channel, providing a rigorous foundation for the conservation of
spatial information under lossless propagation. At the \textit{physical level} (Section~\ref{sec:Line_source}), this geometric structure manifests as \textit{Spatial Information Flows} (\SIFs) defined as the spatial loci along which the spatial coherence (equivalently, the mutual information) decays at the minimum possible rate.
In 2D geometry and asymptotic conditions they
coincide with the optimal sampling curves of Bucci
et al.~\cite{bucci1998representation}.

These results establish that
the \NDF, the \'etendue, the shadow area, the phase-space volume,  the \SIFs, and Bucci's sampling curves are
all manifestations of the same underlying symplectic geometry, and
suggest a path toward a geometric theory of electromagnetic
information.

Conclusions are summarized in Section~\ref{sec:Conclusions}. The
analysis is limited to scalar free-space propagation for clarity of
exposition. The extension to full vector fields, via cross-spectral
density tensors~\cite[Sec.~6.6]{mandel1995optical}, is provided in
the Appendix, where a rigorous proof of the equivalence between
coherent modes and singular functions is given in the full dyadic
setting. A self-contained introduction to Hamiltonian mechanics,
symplectic geometry, and the phase-space framework for antenna
engineers is provided in the Supplementary Material~(SM).

\begin{figure}
	\centering
	\includegraphics[width=0.7\linewidth]{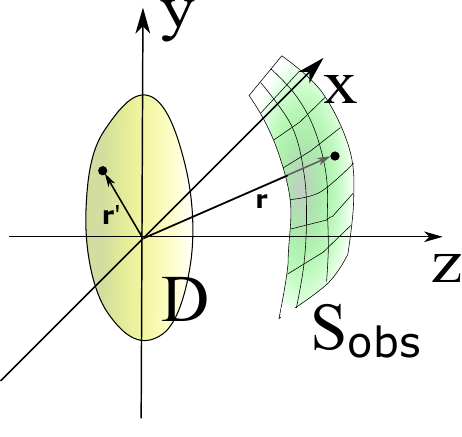}
	\caption{Geometry of the problem for a non-stochastic
		(deterministic) source. The source point $\mathbf{r}'$ belongs
		to the source domain $D$, which may be a volume, a 2D surface,
		or a 1D curve. The observation point $\mathbf{r}$ lies on the
		observation manifold $S_{\mathrm{obs}}$.}
	\label{fig:schema2esitbohrv1}
\end{figure}

\begin{figure}
	\centering
	\includegraphics[width=0.7\linewidth]{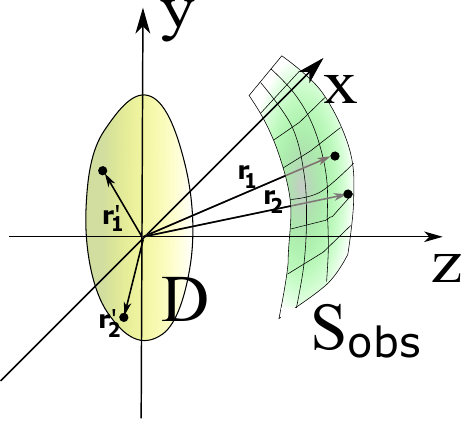}
	\caption{Geometry for a stochastic source. Source points
		$\mathbf{r}'_1$, $\mathbf{r}'_2 \in D$; observation points
		$\mathbf{r}_1$, $\mathbf{r}_2 \in S_{\mathrm{obs}}$.}
	\label{fig:schema2esitbohrv2}
\end{figure}

\section{Background on the Non-Stochastic Approach to EIT}
\label{sec:non_stocastic}

As a preliminary step, let us consider the ``pure spatial communication system'' model shown in Fig.~\ref{fig:schema2esitbohrv1}, consisting of a harmonic electromagnetic source domain $D$ and an observation surface $S_{\mathrm{obs}}$~\cite{migliore2006role}.

With $E^d(\br)$ and $J^d(\br)$ we denote the phasor representation of the electric field and the current density distribution on the source, respectively. The superscript $d$ stands for deterministic. The time dependence $e^{j \bar{\omega} t}$ will be understood and dropped.

Consider the electromagnetic field radiated by a scalar current distribution. Let $\mathcal{H}_J = L^2(D,d\mu_D)$ be a Hilbert space\footnote{The choice of norm depends on the specific problem; $L^2$ spaces are standard in EIT and are adopted throughout this work.} representing the admissible scalar current distributions $J^d(\mathbf{r}'),$ defined on the source domain $D$, and let $\mathcal{H}_E = L^2(S_{\mathrm{obs}}, d\mu_{S_{\mathrm{obs}}})$ represent the observable scalar fields $E^d(\mathbf{r})$ on the observation domain.\footnote{The measure $d\mu_D(\mathbf{r}')$ is $dV'$, $dS'$, or $dl'$ for volumetric, surface, or curve sources. For surface sources, Sobolev spaces such as $H^{-1/2}(D)$ may be more appropriate; the Hilbert--Schmidt analysis below holds for any choice that makes the radiation operator compact.}

The direct relationship between a current distribution $J^d(\mathbf{r}')$ and the resulting radiated electric field $E^d(\mathbf{r})$ is established by the radiation operator given by the integral transformation~\cite{migliore2006role}
\begin{equation}
	E^d(\mathbf{r})
	= \int_D J^d(\mathbf{r}')\, G(\mathbf{r}, \mathbf{r}') \, d\mu_D(\mathbf{r}'),
	\label{eq:RadOp}
\end{equation}
where $G(\mathbf{r}, \mathbf{r}')$ is the free-space Green's function,
\[
G(\mathbf{r},\mathbf{r}')
= -\,j \,\bar\omega \mu_{0} \frac{e^{-j\beta R}}{4\pi R},
\qquad
R = \|\mathbf{r}-\mathbf{r}'\|,
\]
with $\mathbf{r}\in S_{\mathrm{obs}}$, $\beta = \bar\omega \sqrt{\epsilon_0 \mu_0}$ the free-space wavenumber, and $\epsilon_0$, $\mu_0$ the permittivity and permeability of free space, respectively.

The radiation operator, being compact\footnote{A compact operator maps every bounded set into a precompact set (i.e., a set whose closure is compact) and hence into a totally bounded set. From a physical viewpoint, propagation acts as a spatial filter: it smooths the source distribution. Consequently, operators that model propagation are typically compact, since they are smoothing. By contrast, operators that map currents and fields on the same surface, such as the EFIE-type boundary integral operators, are in general not compact. These operators are used to identify the current distribution on the source from a primary excitation (impressed voltage/current or incident field). In this paper, we assume that the currents on the source are known and physically consistent, and that the field is observed on a bounded manifold located at a distance of several wavelengths from the source. Under these conditions, only compact propagation operators are involved.}, can be expanded using Hilbert--Schmidt representation~\cite{kolmogorov1999elements}, obtaining a proper basis for the currents and the field.
\begin{align}
	E^d(\mathbf{r}) &= \sum_n y^d_n {u}_n(\mathbf{r}), \label{eq:E_serie} \\
	J^d(\mathbf{r}') &= \sum_n x^d_n {v}_n(\mathbf{r}')
	\label{eq:EJ_serie}
\end{align}

These bases diagonalize the radiation operator, leading to the relation
\begin{equation}
	y^d_n = \sigma_n x^d_n,
	\label{eq:ysx}
\end{equation}
where $u_n$, $v_n$ are the left and right singular functions and $\sigma_1\ge\sigma_2\ge\dots\to 0$ are the singular values~\cite{kolmogorov1999elements}.

The number of significant singular values determines the number of degrees of freedom of the field~\cite{bucci1989degrees}; the selection of the truncation threshold is discussed later in this section. The singular functions $v_n(\br')$ and $u_n(\br)$ are obtained as solutions of eigenvalue problems~\cite{migliore2021world}. In particular, $u_n$ are the eigenfunctions of the eigenproblem
\begin{equation}
	\int_{S_{\mathrm{obs}}} K_{u}(\mathbf{r}_1, \mathbf{r}_2) \, u_n(\mathbf{r}_2) \, d\mu_{S_{\mathrm{obs}}}(\mathbf{r}_2) = \lambda^d_n \, u_n(\mathbf{r}_1)
	\label{eq:Ku}
\end{equation}
where $\mathbf{r}_1 \in S_{\mathrm{obs}}$, $\lambda^d_n=\sigma_n^2$, and the kernel is:
\begin{equation}
	K_{u}(\mathbf{r}_1, \mathbf{r}_2) = \int_{D} G(\mathbf{r}_1, \mathbf{r}') G^*(\mathbf{r}_2, \mathbf{r}') \, d\mu_D(\mathbf{r}')
	\label{eq:GGc}
\end{equation}
where $G^*(\mathbf{r}_2, \mathbf{r}')$ is the complex conjugate of the Green's function $G(\mathbf{r}_2, \mathbf{r}')$. The $n$-th eigenvalue of this eigenproblem is equal to the square of the $n$-th singular value of the Hilbert--Schmidt decomposition of the radiation operator.

The asymptotic decay rate of the singular values depends on the regularity properties of the kernel of the radiation operator. When the source domain $D$ and the observation domain $S_{\mathrm{obs}}$ are disjoint and bounded, the free-space radiation kernel $G(\mathbf{r},\mathbf{r}')$ is analytic in both variables on $S_{\mathrm{obs}}\times D$. For compact operators with analytic kernels on bounded domains the singular values decay at least exponentially fast~\cite{hille1928characteristic, little1984eigenvalues, konig2013eigenvalue, weyl1911asymptotische, landau1975szego}. In other words, after an initial set of relatively large singular values, one observes a sharp drop-off.

The sharpness and threshold location of this drop-off are controlled by the electrical size of the radiating structure: as the electrical dimension (measured by $\beta$ times a characteristic length of $D$) increases, more singular values remain significant before the rapid decay sets in.

In the non-asymptotic regime, the exact value of the \NDF\ naturally depends on a prescribed accuracy~$\varepsilon>0$, and it is more rigorous to introduce an $\varepsilon$-dependent quantity, \NDFeps. In the electromagnetic literature~\cite{bucci1989degrees, bucci1987spatial}, \NDFeps\ is typically defined in the context of approximation theory: one asks for the minimal number of basis functions needed to approximate any field with error not exceeding $\varepsilon$ in an appropriate norm. In this framework, \NDFeps\ is equal to the number of singular values of the radiation operator whose magnitude is greater than $\varepsilon$, plus one.

Some care is required when translating these approximation-theoretic notions into an information-theoretic setting. Indeed, the \NDFeps\ evaluated in the context of approximation theory typically constitutes an upper bound for the \NDFeps\ estimated in information theory~\cite{migliore2006role}. Informally, approximation theory counts singular functions needed to reduce the residual error below $\varepsilon$, while information theory retains only those whose singular values exceed the noise level~$\varepsilon$. A detailed discussion on this point is reported in Section~\ref{app:functional_models}.

	\section{Information Carried by Stochastic Fields}
	\label{sec:EIT_modelA}
	
	Now, with reference to the same source and receiver domain geometry
	considered in the previous section, let us consider a statistical
	ensemble of current distributions $\{j^{(r)}(\mathbf{r}',t)\}$
	modeled as wide-sense stationary (WSS) stochastic processes. The
	complex analytic representations of field and current at angular
	frequency $\bar\omega$ are $E(\mathbf{r})$ and $J(\mathbf{r}')$
	respectively~\cite{mandel1995optical}. The cross-spectral density (CSD)
	is (Fig.~\ref{fig:schema2esitbohrv2}):
	\begin{align}
		W(\mathbf{r}_1, \mathbf{r}_2)
		&= \mathbb{E}\{E(\mathbf{r}_1)E^*(\mathbf{r}_2)\},
		\label{eq:CSD_field}\\
		W_J(\mathbf{r}'_1,\mathbf{r}'_2)
		&= \mathbb{E}\{J(\mathbf{r}'_1)J^*(\mathbf{r}'_2)\}.
		\label{eq:CSD_source}
	\end{align}
	The CSD acts as the kernel of the Karhunen--Lo\`eve expansion
	(KLE)~\cite{mandel1995optical}:
	\begin{align}
		\int_{S_{\mathrm{obs}}} W(\mathbf{r}_1,\mathbf{r}_2)\,
		\psi_n(\mathbf{r}_2)\,d\mu_{S_{\mathrm{obs}}}(\mathbf{r}_2)
		&= \lambda_n\,\psi_n(\mathbf{r}_1),
		\label{eq:EigWE}\\
		\int_D W_J(\mathbf{r}'_1,\mathbf{r}'_2)\,
		\psi_{Jn}(\mathbf{r}'_2)\,d\mu_D(\mathbf{r}'_2)
		&= \lambda_{Jn}\,\psi_{Jn}(\mathbf{r}'_1).
		\label{eq:WJ}
	\end{align}
	
	The eigenfunctions $\psi_n(\mathbf{r})$ and $\psi_{J_n}(\mathbf{r}')$ represent the completely coherent modes of the field and current, respectively. The field and current ensembles can then be written as a summation over these orthogonal modes:
	\begin{align}
		E(\mathbf{r}) &= \sum_n a_{n} \psi_{n}(\mathbf{r}), \\
		J(\mathbf{r}') &= \sum_n a_{Jn} \psi_{Jn}(\mathbf{r}'),
	\end{align}
	where the expansion coefficients are uncorrelated, satisfying $\mathbb{E} \{ a_{n} a_{m}^* \} = \lambda_{n} \delta_{nm}$ (with $\delta_{nm}$ being the Kronecker symbol). Furthermore, the functions $\psi_n(\mathbf{r})$ and the CSD $W(\mathbf{r}_1, \mathbf{r}_2)$ satisfy the homogeneous Helmholtz equations, ensuring that the CSD ``propagates'' in space according to Maxwell's equations:
	\begin{align}
		\nabla^2_1 W(\mathbf{r}_1, \mathbf{r}_2) + \beta^2 W(\mathbf{r}_1, \mathbf{r}_2) = 0, \\
		\nabla^2_2 W(\mathbf{r}_1, \mathbf{r}_2) + \beta^2 W(\mathbf{r}_1, \mathbf{r}_2) = 0.
	\end{align}
	
	Finally, the degree of spatial coherence is quantified by the spectral degree of coherence, $\mu(\mathbf{r}_1; \mathbf{r}_2)$, which normalizes the CSD by the spectral density (or intensity), $S_E(\mathbf{r}) = W(\mathbf{r},\mathbf{r})$:
	\begin{equation}
		\mu(\br_1;\br_2)
		= \frac{W(\mathbf{r}_1,\mathbf{r}_2)}
		{\sqrt{W(\mathbf{r}_1,\mathbf{r}_1)\,W(\mathbf{r}_2,\mathbf{r}_2)}}.
		\label{eq:SDC}
	\end{equation}
	
	Determining mutual information generally requires knowledge of the full joint probability distribution of the variables, which implies knowing the complete statistics of the underlying process. However, in the case of zero-mean stationary complex Gaussian circularly symmetric processes\footnote{The assumption of a Gaussian distribution is widely accepted in wireless communication systems, as it maximizes Shannon entropy and information content of the field. A circularly symmetric complex Gaussian process is characterized by a probability distribution that is invariant under phase rotations. This assumption implies that the pseudo-covariance matrix (which describes correlations between non-conjugate complex variables) is identically null, thereby simplifying the process's characterization significantly. If the process has a non-zero mean, the CSD used for mutual information calculation must be evaluated on the fluctuating part of the stochastic process. That is, instead of $E(\mathbf{r}_1)$ and $E(\mathbf{r}_2)$, the CSD is evaluated on $E'(\mathbf{r}_1)=E(\mathbf{r}_1)-\mathbb{E}\{E(\mathbf{r}_1)\}$ and $E'(\mathbf{r}_2)=E(\mathbf{r}_2)-\mathbb{E}\{E(\mathbf{r}_2)\}$ \cite{TseVis:B05}.} the statistical properties are entirely described by covariance (or correlation) function, and the mutual information \( I_{MI}(\br_1; \br_2) \) between the two points is given by~\cite{gallager1968information}:
	\begin{equation}
		I_{MI}(\br_1; \br_2)
		= \log_2\left(\frac{1}{1 -|\mu(\br_1; \br_2)|^2}\right)
		\label{eq:MutalInf}
	\end{equation}
	
	This formula quantifies how much information about the field at $\mathbf{r}_1$ can be inferred from the field at $\mathbf{r}_2$\footnote{As $\mathbf{r}_2 \rightarrow \mathbf{r}_1$, $|\mu| \rightarrow 1$ and $I_{MI}$ diverges, reflecting the singular behavior of the noiseless limit. In practice, measurement noise bounds the effective coherence below unity, thereby ensuring that $I_{MI}$ remains finite.}: the lower the mutual information, the more independent information can be encoded at the two points.
	
	Eqs.~\eqref{eq:SDC} and \eqref{eq:MutalInf} allow obtaining the mutual information from the knowledge of CSD. In the following, we will focus our attention on the estimation of CSD.
	
	The CSD of the scalar field, $W(\mathbf{r}_1, \mathbf{r}_2)$, observed at two points $\mathbf{r}_1, \mathbf{r}_2 \in S_{\mathrm{obs}}$, is related to the CSD of the scalar current distribution, $W_J(\mathbf{r}'_1, \mathbf{r}'_2)$, for sources located at $\mathbf{r}'_1, \mathbf{r}'_2 \in D$, by the following integral relationship:
	\begin{equation} \label{eq:Wdi_mu}
		\begin{split}
			W(\mathbf{r}_1, \mathbf{r}_2) = & \int_D \int_{D} W_J(\mathbf{r}'_1, \mathbf{r}'_2) G(\mathbf{r}_1, \mathbf{r}'_1) G^*(\mathbf{r}_2, \mathbf{r}'_2) \\
			& \qquad \times d\mu_D(\mathbf{r}'_1) \, d\mu_D(\mathbf{r}'_2)
		\end{split}
	\end{equation}
	
	In the following analysis, we will assume a spatially uncorrelated source\footnote{Rigorously speaking, a source cannot be perfectly spatially uncorrelated \cite{plonus1991spatial}. However, the correlation is subject to the low-pass filtering process of the free-space Green's function that tends to filter the higher spatial frequencies \cite[Sec.~5.3]{mandel1995optical}. This implies that the approximation of a spatially incoherent source does not significantly affect the field results when the source's intrinsic spatial correlation length is sufficiently short (let us say, shorter than $\lambda/2$), as propagation naturally increases the coherence distance.}, which can be considered the 'richest source' in terms of spatial information. For such a source, the CSD of the current is given by $W_{J}(\br'_1, \br'_2) = S_J(\mathbf{r}'_1) \delta(\br'_1 - \br'_2)$, where $S_J(\mathbf{r}'_1) = W_J(\br', \br')$ is the source spectral density per unit angular frequency.
	
	Under this hypothesis, the formulas \eqref{eq:Wdi_mu} simplify as
	\begin{equation}
		W(\mathbf{r}_1,\mathbf{r}_2)=\int_D S_J(\mathbf{r}')\, G(\mathbf{r}_1,\mathbf{r}')\, G^*(\mathbf{r}_2,\mathbf{r}')\, d\mu_D(\mathbf{r}')
		\label{eq:Wx}
	\end{equation}
	
	This integral is of the same form as is encountered in the evaluation of the field diffracted by an aperture $D$ in an opaque screen at a point $\br_1$ illuminated by a harmonic spherical wave centered in $\br_2$, or also of the scattered field in Born approximation. This similarity is at the basis of many properties that the CSD shares with the theory of diffraction and scattering integrals\footnote{For a planar source and under far-field conditions, Eq.~\eqref{eq:Wx} has the same structure as the Van Cittert--Zernike theorem used in optics.}.
	
	The \NDF\ of a stochastic process can be obtained by its Karhunen--Lo\`eve expansion. With reference to Eq.~\eqref{eq:EigWE}, we recall that the eigenvalues $\lambda_{n}$ are real and non-negative, and exhibit a step-like behavior: a finite number of them are significant (close to a maximum value), while the rest are negligible. The number of these significant modes is the \NDF\ of the stochastic field.
	
	It can be easily shown that the \NDF\ of the field radiated by an uncorrelated source is identical to that of an auxiliary deterministic problem, described by an effective operator with the kernel $G'(\mathbf{r}, \mathbf{r}') = G(\mathbf{r}, \mathbf{r}') \sqrt{S_J(\mathbf{r}')}$.
	
	Let us consider the eigenproblems in Eqs.~\eqref{eq:Ku} and \eqref{eq:EigWE}, whose kernels are given by Eqs.~\eqref{eq:GGc} and \eqref{eq:Wx}, respectively. We note that, for a spatially uncorrelated homogeneous source\footnote{A homogeneous source is a source whose statistical properties are uniform across its surface, i.e., whose spectral density is constant on the source. A spatially uncorrelated homogeneous source is an idealization. Physically realizable sources are only partially coherent, i.e., \emph{quasi-homogeneous} with a small but non-zero coherence area.} with $S_J(\mathbf{r}') = 1$, the two kernels coincide, i.e., $W(\mathbf{r}_1,\mathbf{r}_2) = K_u(\mathbf{r}_1,\mathbf{r}_2)$. The eigenvalue problem~\eqref{eq:EigWE} is therefore identical to~\eqref{eq:Ku}. It follows that the eigenfunctions of the CSD operator are identical to the left singular functions of the radiation operator, i.e. $\psi_n(\mathbf{r}) = u_n(\mathbf{r})$, and that the corresponding eigenvalues also coincide, namely $\lambda_n = \sigma_n^{2}$. As a result, \textit{the stochastic \NDF\ coincides with the deterministic \NDF}, and the same equivalence holds for any $\varepsilon$-truncated measure of the number of degrees of freedom (i.e. \NDFepsv, \cite{migliore2008electromagnetics}) based on the decay of the eigenvalue sequence.
	
	This key equivalence---valid also for full vector fields as proved in Appendix~\ref{app:theorem}---explains the remarkable success of classical electromagnetic methods in the EIT analysis of spatial information. Its deeper mathematical foundation is established in the following section through functional-set analysis.
	
	\begin{figure}[t]
		\centering
		\includegraphics[width=1.0\linewidth]{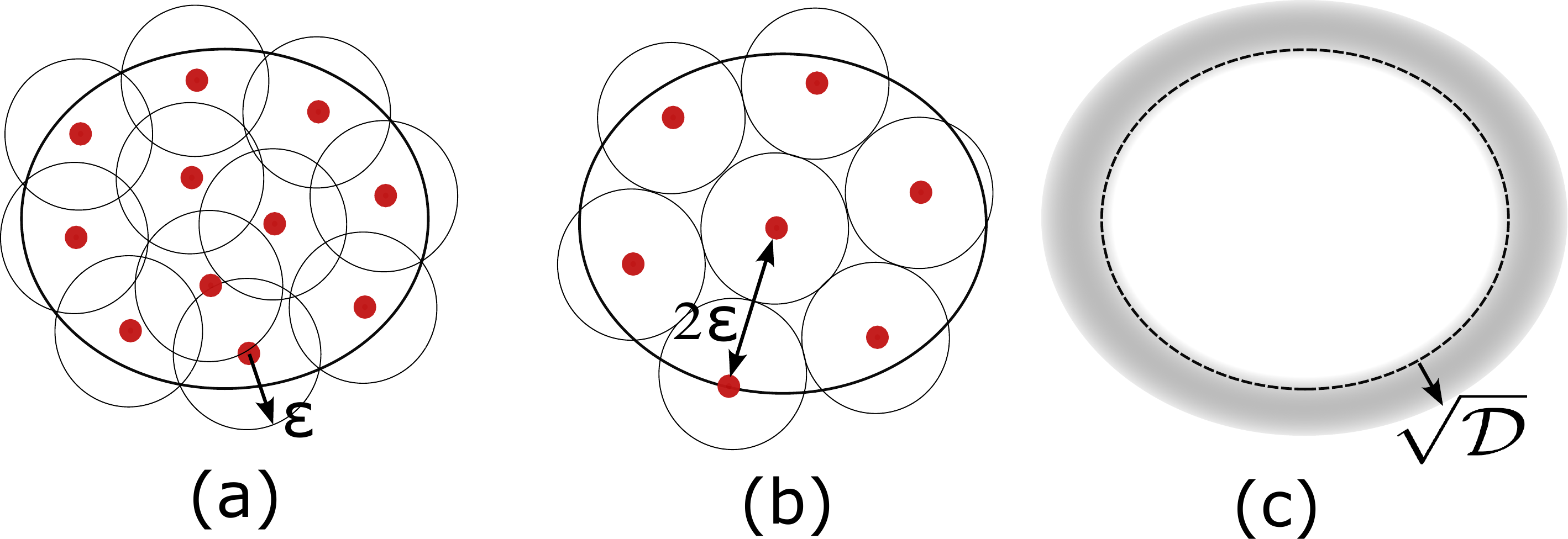}
		\caption{Geometric interpretation of deterministic and stochastic information measures. (a) $\varepsilon$-covering of $Y_{\det}$. (b) $\varepsilon$-packing of $Y_{\det}$. (c) Stochastic formulation: sphere hardening makes the rate--distortion problem asymptotically equivalent to deterministic sphere packing, both quantifying the same symplectic volume $G_e/\lambda^2$.}
		\label{fig:ballsdetprobv2}
	\end{figure}
	
	\section{A Unified View Based on Functional Sets}
	\label{app:functional_models}
	
	The equivalence established in Section~\ref{sec:EIT_modelA} between
	deterministic and stochastic \NDF\ has a precise mathematical
	explanation in terms of functional sets. This analysis, based on
	Kolmogorov's $\varepsilon$-capacity
	theory~\cite{tikhomirov1993varepsilon}, reveals the common geometric
	skeleton underlying both formulations and resolves the apparent
	multiplicity of \NDF\ definitions.

The functional-set approach in EIT was proposed for the first time in \cite{migliore2008electromagnetics} and represents an application of Kolmogorov's theory of $\varepsilon$-capacity \cite{tikhomirov1993varepsilon} to electromagnetic problems. It was successively extended \cite{migliore2018horse, franceschetti2017wave, lim2017information} and further discussed in \cite[Appendix~I]{migliore2021world}, although the analysis in these works was limited to non-stochastic electromagnetic models.

Results on Kolmogorov’s $\varepsilon$-entropy theory applied to stochastic fields are relatively scarce. The emergence of a finite-dimensional signal subspace due to spectral concentration was first exemplified in the seminal work of Slepian, Pollak, and Landau on time- and band-limited functions~\cite{slepian1962prolateIII}, while an analysis of the $\varepsilon$-entropy of stochastic processes was given in~\cite{posner1967epsilon}. 

In this section, we extend the functional-set approach to stochastic electromagnetic fields in order to obtain a unified view of deterministic and stochastic EIT. To this end, we exploit the concentration property of Gaussian fields---namely the Sphere Hardening Theorem\footnote{The sphere hardening phenomenon in information theory states that realizations of high-dimensional Gaussian processes concentrate on a thin shell with overwhelming probability, rendering probabilistic soft ``spheres'' asymptotically equivalent to deterministic hard spheres.
	~\cite{gallager1968information,pinsker1964information,berger1971rate}. This property is the probabilistic analogue of the geometric concentration of volume near the surface of high-dimensional spheres in Euclidean spaces}~\cite{gallager1968information}.

Starting with the deterministic model, each current distribution $J^d(\mathbf{r}')$ can 
be represented as a point in the corresponding Hilbert space $\mathcal{H}_J$. Let 
$X_{det}$ denote the set of all admissible source currents. We assume that the source 
energy is bounded, such that $X_{det}$ corresponds to the unit ball in $\mathcal{H}_J$ 
($\lVert J^d \rVert \le 1$).

Analogously, each radiated field configuration $E^d(\mathbf{r})$ is a point in the 
observation Hilbert space $\mathcal{H}_E$. The radiation operator maps the unit ball 
$X_{det}$ into a hyperellipsoid $Y_{det}$ within the observation space 
(see Eq.~\eqref{eq:ysx}). The geometry of $Y_{det}$ is strictly defined by the operator's 
spectral properties, where the length of the $n$-th semi-axis equals the singular value 
$\sigma_n$, consistently with classical results on compact operators in Hilbert spaces 
\cite{pinkus2012n}.

Given a maximum acceptable approximation error $\varepsilon$, approximation theory is 
naturally formulated in terms of the \emph{covering number} $N(\varepsilon)$, defined as 
the minimal number of open balls of radius $\varepsilon$ needed to cover $Y_{det}$ 
\cite{tikhomirov1993varepsilon} (see Fig.~\ref{fig:ballsdetprobv2}(a)). Conversely, 
information theory relies on the \emph{packing number} $M(\varepsilon)$, defined as the 
maximal number of disjoint open balls of radius $\varepsilon/2$ that can be placed within 
$Y_{det}$ (equivalent to the maximum number of $\varepsilon$-separated points) 
\cite{tikhomirov1993varepsilon} (see Fig.~\ref{fig:ballsdetprobv2}(b)).

These quantities satisfy the fundamental inequality \cite{tikhomirov1993varepsilon}:
\begin{equation}
	M(2\varepsilon) \le N(\varepsilon) \le M(\varepsilon).
\end{equation}
Defining the effective dimension $d_s$ (the NDF) via the Kolmogorov limit 
$d_s = \lim_{\varepsilon\to 0} \frac{\log N(\varepsilon)}{\log(1/\varepsilon)}$, and 
applying the inequality above, it follows that
\begin{equation}
	\lim_{\varepsilon\to 0} \frac{\log M(2\varepsilon)}{\log(1/\varepsilon)} \le d_s \le 
	\lim_{\varepsilon\to 0} \frac{\log M(\varepsilon)}{\log(1/\varepsilon)}.
\end{equation}
Since the scale factor of $2$ vanishes in the logarithmic limit, both covering and packing 
numbers yield the same effective dimension $d_s$. While the effective dimensions derived 
from these two viewpoints may differ slightly in the non-asymptotic regime, they converge 
to the intrinsic NDF for a large number of degrees of freedom, in agreement with general 
results in approximation theory \cite{pinkus2012n}.

We now consider the stochastic approach. Each realization of the stochastic 
electromagnetic field is a functional point in $\mathcal{H}_E$. For a Gaussian random 
field, the set of realizations occurring with high probability $(1-\delta)$ forms an 
effective signal set $Y_{stoch}$, which is a hyperellipsoid in the functional space 
(see Fig.~\ref{fig:ballsdetprobv2}(c)). Its geometry is defined by the spectral properties 
of the covariance operator through the Karhunen--Lo\`eve expansion, as established in 
classical information-theoretic treatments of random processes 
\cite{gallager1968information, berger1971rate}.

Rate--Distortion Theory \cite{gallager1968information, berger1971rate} provides a natural
measure of complexity for $Y_{\mathrm{stoch}}$. For a Gaussian source with
Karhunen--Loève eigenvalues $\{\lambda_n\}$ and mean-square distortion
constraint $\mathcal{D}$, the minimum number $N$ of statistically independent
modes retained is defined by $\sum_{n=N+1}^{\infty} \lambda_n \le \mathcal{D}$.
This admits a direct geometric interpretation in terms of sphere packing,
consistent with the classical treatment of rate--distortion in Hilbert spaces
\cite{pinsker1964information}. Consequently, the stochastic NDF corresponds to
the number of effective dimensions exceeding the noise threshold $\mathcal{D}$.

To establish rigorous geometric equivalence between $Y_{det}$ and $Y_{stoch}$, the boundary conditions defining the signal ellipsoids must first coincide through a condition of \textit{source equivalence}. In the deterministic model, the source is defined as a unit-energy ball ($\lVert J^d \rVert \le 1$); for the stochastic ellipsoid to match this geometry, the source must be modeled as a spatially incoherent white process with unit spectral density, $S_J(\mathbf{r}) = 1$. Under this hypothesis, the source covariance reduces to the identity, implying that the eigenvalues $\lambda_n$ are equal to the squared singular values $\sigma_n^{2}$, as in classical Gaussian source models \cite{pinsker1964information}.

With the signal space geometries thus aligned, the equivalence of the information metrics relies on the asymptotic behavior of the stochastic error. The validity of treating the probabilistic set $Y_{stoch}$ as a rigid geometric body is justified in the asymptotic regime by the phenomenon of \textit{sphere hardening} \cite{gallager1968information}, which renders the Rate--Distortion problem isomorphic to the deterministic sphere-packing problem. This asymptotic rigidity further necessitates a \textit{threshold equivalence}, where the accuracy metrics define the same sphere of uncertainty by requiring $\varepsilon = \sqrt{\mathcal{D}}$. Under this unified identification, the stochastic Rate--Distortion function corresponds geometrically to the deterministic packing number $M(2\varepsilon)$.

In conclusion, in the asymptotic limit, the stochastic and deterministic formulations converge as statistical fluctuations vanish through the concentration of measure. This suggests that the identical results obtained across distinct frameworks are not coincidental, but rather manifestations of a deeper "geometric skeleton" inherent to the electromagnetic field.
The remainder of this paper is dedicated to identifying this foundational structure and investigating its physical implications using symplectic geometry and phase-space analysis.
	

\section{The Role of Information and Power Transfer in the \NDF}
\label{sec:EIT_modelB}

The functional-set analysis of the previous section showed that all
\NDF\ definitions converge to the same geometric skeleton: the
hyperellipsoid $Y_{\mathrm{det}}$ in the observation Hilbert space,
whose semi-axes are the singular values of the radiation operator.
The question that remains is what determines the effective dimension
of this hyperellipsoid. We now address this question in the
asymptotic regime, showing that the dimension is governed by the
geometric \'etendue of the source--observer configuration.

The connection is established through a phase-space representation
of the power cross-spectral density. The natural mathematical
framework for this transition is symplectic
geometry~\cite{frankel2004geometry, zworski2012semiclassical}---the
intrinsic geometry of Hamiltonian mechanics and, in the
semiclassical limit, of electromagnetic ray
propagation.\footnote{Symplectic geometry is the branch of differential
	geometry that studies manifolds equipped with a closed,
	non-degenerate 2-form $\varpi$. A \emph{symplectic
		transformation} is one that preserves $\varpi$---or
	equivalently, any transformation generated by a Hamiltonian
	function via Hamilton's equations, since Hamiltonian flows
	preserve $\varpi$ by construction. 
	Free-space propagation, reflection by perfect conductors, and refraction by lossless dielectrics are all symplectic transformations and conserve the symplectic phase-space volume $d^2q\,d^2k_{\parallel}$ on the cotangent bundle $T^*M$ of the surface M. 
	In the presence of losses, by contrast, the canonical
	momentum acquires an imaginary part and the Hamiltonian
	vector field has non-zero divergence: phase-space volume
	contracts, the symplectic structure is broken, and
	Liouville's theorem no longer holds. The symplectic
	framework of this paper therefore applies to lossless
	propagation. A simple self-contained introduction is provided in SM, Section~I.}
For clarity, the analysis below
relies primarily on basic tools of differential
geometry~\cite{docarmo1992riemannian, frankel2004geometry}; the
symplectic structure is identified progressively as the analysis
unfolds, after the relevant geometric concepts have been
introduced.

To extend the framework to non-planar geometries, we
consider a field on a smooth $n$-dimensional manifold $M$
embedded in $\mathbb{R}^m$, with field correlation length
$L_c \ll R_c$, where $R_c$ is the local radius of
curvature. Under this condition the field is locally
well-approximated by plane waves over the extent of a
correlation patch, justifying the local application of
the Wigner transform. 
At each point $\mathbf{r}\in M$, the phase-space structure requires two dual spaces: the \textit{tangent space} $T_{\mathbf{r}}M$, containing local displacement vectors $\Delta\mathbf{r}$ and equipped with the covariant metric $g_{ij}$, and the \textit{cotangent space} $T^*_{\mathbf{r}}M$, equipped with the dual metric $g^{ij}=(g_{ij})^{-1}$,\footnote{The dual metric satisfies $g_{ij}g^{jl}=\delta_i^l$: while $g_{ij}$ measures distances between neighbouring points, $g^{ij}$ measures distances between neighbouring wavevectors.} whose natural elements are tangential covectors~---in particular, the components of the propagation wavevector $\mathbf{k}=\beta\hat{\mathbf{s}}$ sensed along the surface at $\mathbf{r}$.

Consequently, the full wavevector $\beta\hat{\mathbf{s}}\in\mathbb{R}^m$ is not an element of $T^*_{\mathbf{r}}M$; only its tangential projection $\mathbf{k}_{\parallel} = \beta\,\Pi_{T_{\mathbf{r}}M}(\hat{\mathbf{s}}) \in T^*_{\mathbf{r}}M$ is the natural cotangent variable,\footnote{Rigorously speaking, elements of $T^*_{\mathbf{r}}M$ are 1-forms, not vectors~\cite{frankel2004geometry}, and this geometric operation is formally defined as the pullback of the ambient wavevector $\mathbf{k}$ onto the cotangent space $T^*_{\mathbf{r}}M$. However, the metric $g_{ij}$ establishes a canonical isomorphism (the `musical isomorphism' $\flat$) between $TM$ and $T^*M$, so that covectors and tangent vectors can be identified whenever the surface is locally flat ($g_{ij} \simeq \delta_{ij}$). Under this condition, which holds throughout this paper by the assumption $L_c \ll R_c$, the pullback reduces to the ordinary dot product along the surface, and the distinction between vectors and covectors is immaterial for all practical computations. Further details are provided in the Supplementary Material, Section~IV.} with magnitude $|\mathbf{k}_{\parallel}| = \beta\sin\theta$. The factor $\cos\theta$ appearing later in the \'etendue integral arises as the Jacobian of the transformation $d^2k_{\parallel} = \beta^2\cos\theta\,d\Omega$. The cotangent bundle $T^*M$---which associates $T^*_{\mathbf{r}}M$ with every point $\mathbf{r}\in M$, globally pairing positions and their conjugate tangential wavevectors---is the natural phase space for the field.\footnote{A \textit{bundle} associates a vector space with every point of a manifold. The tangent bundle $TM$ associates $T_{\mathbf{r}}M$ (displacement vectors) with each point, while the cotangent bundle $T^*M$ associates $T^*_{\mathbf{r}}M$ (tangential wavevectors). Together they form the complete phase-space description. A self-contained introduction, starting from the familiar planar aperture, is provided in SM, Sections~IV and~V.}


The analysis is formulated in terms of the power cross-spectral
density $W_p(\mathbf{r}_1,\mathbf{r}_2)$, defined as the
cross-spectral density of the time-averaged power flux through
$M$.\footnote{Here $W_p$ denotes the field cross-spectral density
	expressed in power-flux units, $W_p \equiv W/2\zeta_0$, where
	$\zeta_0=\sqrt{\mu_0/\epsilon_0}$ is the free-space wave
	impedance; on the diagonal $\mathbf{r}_1=\mathbf{r}_2$ it reduces
	to the time-averaged power density. This identification holds when
	two conditions are simultaneously satisfied. First, the
	observation distance must satisfy $R\gg\lambda$, so that
	evanescent modes are negligible and all field components are
	propagating plane waves with $|\mathbf{k}|=\beta$. Second, the
	source must be spatially incoherent or only partially coherent:
	for a source with finite coherence length $L_c$, the cross terms
	between different plane-wave components $(\mathbf{k},\mathbf{k}')$
	in the $\mathbf{E}\times\mathbf{H}^*$ product cancel by
	statistical averaging, so that the power density reduces to a
	single integral over $\mathbf{k}$ proportional to $W$. For the
	spatially incoherent source of Section~\ref{sec:Line_source}
	($S_J=\mathrm{const}$), both conditions are satisfied in the
	radiation zone and the relation becomes exact as $R\to\infty$.
	For a perfectly coherent source ($L_c\to\infty$), the cross terms
	do not cancel and the relation does not hold in general. In the
	general case, $W_p$ must be obtained from the cross-correlation of
	the electric and magnetic fields via the Poynting
	theorem~\cite[Sec.~6.6]{mandel1995optical}.}.
The use of a two-point spatial correlation is not merely a
statistical convenience, but a geometric necessity. In the
harmonic domain, the canonical momentum of the electromagnetic
field becomes algebraically locked to its amplitude, causing
the temporal phase space to collapse into a degenerate
line---a conceptual pitfall we term the ``phasor trap''
(discussed in detail in Section~II of the Supplementary
Material). To escape this trap and restore a valid,
non-degenerate phase space (the \textit{symplectic} structure
of which is formalized later in this section), one must
transition from a single-point amplitude description to a
two-point spatial correlation.

The bridge between the stochastic description and the restored spatial
phase space is provided by the Wigner Distribution Function
(WDF)~\cite{bastiaans2009wigner, gradoni2014wigner}, a
quasi-probability distribution\footnote{The WDF is termed
	quasi-probability because, unlike a true probability density, it
	can take negative values in regions of strong wave interference.
	In the asymptotic (geometrical-optics) limit, these negative
	regions vanish and the WDF converges to the non-negative classical
	radiance~\cite{mandel1995optical}.} that jointly encodes the
spatial support of the field and its directional content. Its
Wigner transform on $M$ is:
\begin{equation}
	\WDFP(\mathbf{r},\mathbf{k})
	= \int_{T_{\mathbf{r}}M}
	W_p\!\!\left(\mathbf{r}+\tfrac{\Delta\mathbf{r}}{2},
	\mathbf{r}-\tfrac{\Delta\mathbf{r}}{2}\right)
	e^{-j\mathbf{k}\cdot\Delta\mathbf{r}}\,d(\Delta\mathbf{r}),
	\label{eq:wigner_def_unified}
\end{equation}
where $\mathbf{r}=(\mathbf{r}_1+\mathbf{r}_2)/2$ is the midpoint
and $\Delta\mathbf{r}=\mathbf{r}_1-\mathbf{r}_2$ is the separation
vector, integrated over the tangent space $T_{\mathbf{r}}M$.
Since $\Delta\mathbf{r}\in T_{\mathbf{r}}M$, the exponent
$\mathbf{k}\cdot\Delta\mathbf{r}$ depends only on the tangential
projection $\mathbf{k}_{\parallel}=\Pi_{T_{\mathbf{r}}M}(\mathbf{k})
\in T^*_{\mathbf{r}}M$; the normal component
$k_n=\mathbf{k}\cdot\hat{\mathbf{n}}$ plays no role. The WDF is
therefore naturally supported on the set of propagating directions:
for a field satisfying the Helmholtz equation in $\mathbb{R}^m$,
this is the forward hemisphere
\begin{equation}
	\Xi(\mathbf{r})
	= \bigl\{\mathbf{k}\in\mathbb{R}^m :
	|\mathbf{k}|=\beta,\;
	\hat{\mathbf{n}}(\mathbf{r})\cdot\hat{\mathbf{k}}>0
	\bigr\},
	\label{eq:Xi_def}
\end{equation}
which will play a central role in the definition of the phase-space
volume below. The WDF thus lives on the cotangent bundle $T^*M$,
with $\mathbf{r}$ and $\mathbf{k}_{\parallel}$ forming the conjugate
pair whose symplectic structure is analysed below.

In geodesic normal coordinates $q=(q^1,\dots,q^n)$ centered at
$\mathbf{r}$, the metric admits the Taylor
expansion\footnote{Throughout this section, repeated indices imply
	summation according to the Einstein convention
	(e.g., $q^k q^l \equiv \sum_{k,l} q^k q^l$)}~\cite{docarmo1992riemannian}:
\begin{equation}
	g_{ij}(q) = \delta_{ij}
	- \tfrac{1}{3}R_{ikjl}(0)\,q^kq^l + O(|q|^3),
	\label{eq:metric_expansion}
\end{equation}
where $R_{ikjl}\sim 1/R_c^2$ is the Riemann curvature tensor.
Under $L_c\ll R_c$ the curvature term is negligible, so
$g_{ij}\simeq\delta_{ij}$ and $M$ is locally indistinguishable from
its tangent space. The dual metric satisfies:
\begin{equation}
	g^{ij}(q) = \delta^{ij}
	+ \tfrac{1}{3}R^{i\phantom{k}j}_{\phantom{i}k\phantom{j}l}(0)
	\,q^kq^l + O(|q|^3),
	\label{eq:dual_metric_expansion}
\end{equation}
with the opposite sign, reflecting that spatial contraction due to
curvature implies spectral expansion in the cotangent fiber.

The fundamental consequence for the phase-space volume element is the
following. The volume elements on the tangent and cotangent fibers are
$d\mu_g = \sqrt{\det g_{ij}}\,d^nq$ and
$d\mu_{g^*} = \sqrt{\det g^{ij}}\,d^nk_{\parallel}$ respectively.
Their product gives:
\begin{equation}
	d\tilde\Omega
	= \sqrt{\det g_{ij}\cdot\det g^{ij}}\,d^nq\,d^nk_{\parallel}
	= d^nq\,d^nk_{\parallel},
	\label{eq:Liouville}
\end{equation}
since $\det g^{ij} = 1/\det g_{ij}$, so the two Jacobians cancel
exactly.

This result, obtained here using basic differential geometry,
has a deeper interpretation in symplectic geometry: it is a
consequence of the canonical symplectic 2-form
$\varpi = \sum_i dk_{\parallel,i}\wedge dq^i$ on
$T^*M$~\cite{frankel2004geometry}, which encodes the pairing between
positions $q^i$ and conjugate wavevectors $k_{\parallel,i}$. Two properties of
$\varpi$ are essential. First, it is \emph{closed} ($d\varpi=0$):
this is the phase-space analogue of a solenoidal field in
electromagnetism---$\varpi$ has no local ``sources'' or ``sinks'' in
phase space, guaranteeing that the pairing between positions and
momenta is globally consistent and conserved along Hamiltonian flows.
Second, it is \emph{non-degenerate}: for every non-zero vector $\delta v$ tangent to the phase space $T^*M$, there exists another tangent vector $\delta u$ such that $\varpi(\delta v, \delta u) \neq 0$.  Physically, this means that every direction in phase space
is ``felt'' by the symplectic structure: there are no hidden or
degenerate directions along which positions and momenta are
decoupled. Together, these two properties imply
that the volume form $\varpi^n/n!$---the $n$-fold product of $\varpi$
with itself\footnote{The exterior product $\varpi^n/n!$ is the
	antisymmetric $n$-fold product of the 2-form $\varpi$ with itself.
	In local coordinates $(q^1,\dots,q^n,k_{\parallel,1},\dots,k_{\parallel,n})$ it reduces
	to $dk_{\parallel,1}\wedge dq^1\wedge\cdots\wedge dk_{\parallel,n}\wedge dq^n$, which is
	simply the ordinary volume element $d^nq\,d^nk_{\parallel}$ on the $2n$-dimensional
	phase space. The key point is that this reduction holds in
	\emph{any} coordinate system, because $\varpi$ is intrinsic.},
which in local coordinates equals the ordinary
phase-space volume element $d^nq\,d^nk_{\parallel}$---is preserved by the
Hamiltonian flow.
This is Liouville's theorem: the phase-space volume element
$d^nq\,d^nk_{\parallel}$ is preserved under the Hamiltonian flow,
independently of the curvature of $M$.

%
The globally accessible phase-space volume is defined by integrating
the WDF over the spectral support $\Xi(\mathbf{r})$ defined
in~\eqref{eq:Xi_def} and over the entire manifold $M$. The spectral
power density at each point is recovered as:
\begin{equation}
	S_p(\mathbf{r}) = \frac{1}{(2\pi)^n}
	\int_{\Xi(\mathbf{r})} \WDFP(\mathbf{r},\mathbf{k})\,d\sigma(\mathbf{k}),
\end{equation}
where $d\sigma(\mathbf{k})$ denotes the projected (cotangent-plane) measure, i.e. the image of the hemisphere measure under the bijection $\hat{\mathbf{s}} \mapsto \mathbf{k}_{\parallel}$, so that $\int_{\Xi(\mathbf{r})}(\cdot)\,d\sigma(\mathbf{k})$ is understood as an integral over the image disk $\{|\mathbf{k}_{\parallel}|<\beta\}$, confirming that
$\WDFP(\mathbf{r},\mathbf{k})$ provides a local spatial-directional
decomposition of the spectral power density.
The same WDF, integrated over the full spectral support, gives the globally accessible
phase-space volume:
\begin{equation}
	\Vol_{\mathrm{PS}}
	\equiv \int_M\int_{\Xi(\mathbf{r})}
	d\sigma(\mathbf{k})\,d\mu_g(\mathbf{r}).
	\label{eq:VolPS}
\end{equation}
The \NDF\ is obtained by counting the number of elementary
phase-space cells fitting into $\Vol_{\mathrm{PS}}$. Each
independent mode occupies a volume $(2\pi)^n$ in phase
space\footnote{This follows directly from the Fourier structure
	of the Wigner transform~\eqref{eq:wigner_def_unified}: the WDF
	is a Fourier transform with respect to the separation variable
	$\Delta\mathbf{r}\in T_{\mathbf{r}}M$. By Parseval's theorem,
	each orthogonal mode occupies a phase-space cell of volume
	$(2\pi)^n$ in the cotangent coordinates $(q^i, k_{\parallel,i})$.
	Each conjugate pair $(q^i, k_{\parallel,i})$ contributes a factor
	$2\pi$, giving the elementary cell
	$\prod_i \Delta q^i \cdot \Delta k_{\parallel,i} = (2\pi)^n$---the
	Heisenberg cell in $n$ dimensions.}.
Since \eqref{eq:Liouville} shows that the size of this cell is
coordinate-invariant and curvature-independent---equal to $\varpi^n/n!$
in symplectic geometry, regardless of the geometry of $M$---the mode
count is~\cite{zworski2012semiclassical}:
\begin{equation}
	\NDF = \frac{\Vol_{\mathrm{PS}}}{(2\pi)^n}.
	\label{eq:NDF}
\end{equation}
This result holds whether $M$ is closed (capturing all radiated
information) or bounded with boundary (intercepting a specific subset
of degrees of freedom).

Two complementary guarantees follow from symplectic geometry.
Liouville's theorem provides a scalar constraint: $\Vol_{\mathrm{PS}}$
is conserved under Hamiltonian propagation, so the \NDF\ is invariant
regardless of the observation surface chosen. Physically, this is the
phase-space counterpart of Poynting's theorem: in a lossless medium,
the total radiated power crossing any closed surface enclosing the
source is the same, and Liouville's theorem guarantees that the count
of independent spatial channels shares this invariance.

Gromov's Non-Squeezing Theorem~\cite{gromov1985pseudo}---the \textit{Symplectic Camel} theorem---provides a structural constraint of a different nature. The theorem states that, in dimensionless phase-space coordinates $\tilde{q}^i = \beta q^i$ and $\tilde{k}_i = k_{\parallel,i}/\beta$, a symplectic ball
\begin{equation}
	B^{2n}(r) = \Bigl\{ (\mathbf{\tilde{q}}, \mathbf{\tilde{k}}) \in \mathbb{R}^{2n} : \sum_{i=1}^n \bigl[ (\tilde{q}^i)^2 + (\tilde{k}_{i})^2 \bigr] \le r^2 \Bigr\} \label{eq:symball}
\end{equation}
cannot be mapped by any Hamiltonian transformation into the cylinder
\begin{equation}
	Z^{2n}(R) = \Bigl\{ (\mathbf{\tilde{q}}, \mathbf{\tilde{k}}) \in \mathbb{R}^{2n} : (\tilde{q}^1)^2 + (\tilde{k}_{1})^2 \le R^2 \Bigr\}, \label{eq:symcyl}
\end{equation}
which is unbounded in all remaining $2n-2$ directions, unless $R\ge r$. In other words, the symplectic cross-section of the ball on \emph{any} conjugate plane $(\tilde{q}^i,\tilde{k}_{i})$ has a minimum area $\pi r^2$ that no Hamiltonian transformation can reduce---for each conjugate pair individually, independently of what happens to the other pairs. While Liouville fixes the total phase-space volume, Gromov establishes that symplectic regions cannot be arbitrarily deformed: the ``symplectic width'' $\pi r^2$ is a rigid geometric invariant.

Applied to the radiation channel, this means
that the phase-space cell occupied by each independent spatial
mode cannot be compressed below its minimum symplectic size
by any Hamiltonian manipulation of the channel---no antenna
array design, beamforming strategy, or signal processing can
circumvent this geometric lower bound, provided the propagation
is lossless and the phase-space regions can be approximated as
compact.\footnote{These two conditions---losslessness and compactness
	of the phase-space support---define the domain of validity of
	Gromov's theorem in the present context. Losslessness ensures
	that the Hamiltonian structure is preserved and the
	phase-space volume is conserved; in the presence of losses
	both Liouville's and Gromov's guarantees cease to hold.
	Compactness requires that each independent spatial mode
	occupies a well-defined, bounded region in phase space, so
	that it can be approximated by a symplectic ball $B^{2n}(r)$
	to which the theorem applies. In the asymptotic regime, where the geometric-optics approximation
	holds, each mode is well localised both spatially (within the
	source region) and spectrally (within the propagating hemisphere
	$\Xi(\mathbf{r})$), and the compactness condition is
	well satisfied. Outside this regime---for example in the
	near-field or for electrically small sources---the modes are
	not well localised in phase space and the theorem provides
	only an idealized lower bound. A discussion is
	provided in SM, Section~III.}

While Liouville guarantees the \textit{count} of independent
channels, Gromov guarantees their \textit{individuality} within
this domain of validity: distinct spatial modes cannot be merged
by any Hamiltonian transformation. The \NDF\ is therefore a
symplectically protected invariant, under the conditions of lossless
propagation in the far-field regime that underlie the present
analysis.

To evaluate the phase-space volume for the two-dimensional
case ($n=2$, $m=3$), we recall that under the local flatness
assumption $g_{ij}\simeq\delta_{ij}$ the Riemannian measure
reduces to $d\mu_g\simeq dS$, and the global integral over
$T^*M$ assembles by patching together local contributions on
each tangent plane.

We also note that from a physical perspective, Poynting's theorem
dictates that in a lossless medium, the total outward flux of real
power is invariant across any closed manifold $M$ enclosing the
source. Consequently, specializing to the case of a 2D surface
embedded in 3D space ($n=2$, $m=3$) and restricting the analysis to
the \NDF\ associated with radiating modes, we can identify the
mathematical manifold $M$ with any closed surface surrounding the
source, provided that it lies entirely in the source-free region
(i.e., outside the sources), including surfaces arbitrarily close to
the physical emitting surface $S$. In the following, $S$ is assumed
to be convex.

Let $(u,v)$ be local coordinates on $S$ near $\mathbf{r}$. The
cotangent coordinates $(k_u,k_v)$ are the components of the
wavevector covector $\mathbf{k}_{\parallel} \in T^*_{\mathbf{r}}S$---the
projections of the propagation vector $\beta\hat{\mathbf{s}}$ onto
the local tangent plane:
\begin{equation}
	(k_u,k_v)
	= \beta\,\Pi_{T_{\mathbf{r}}S}(\hat{\mathbf{s}})
	= \beta(\sin\theta\cos\phi,\,\sin\theta\sin\phi),
	\label{eq:cotangent_coords}
\end{equation}
where $(\theta,\phi)$ are the polar angles of $\hat{\mathbf{s}}$
measured from the outward normal $\hat{\mathbf{n}}(\mathbf{r})$.
This projection is one-to-one only for the outward-propagating
hemisphere $\Omega^+(\mathbf{r}) = \{\hat{\mathbf{s}}:
\hat{\mathbf{n}}(\mathbf{r})\cdot\hat{\mathbf{s}}>0\}$.
The Jacobian of the bijection $\hat{\mathbf{s}}\mapsto(k_u,k_v)$
from this hemisphere to the cotangent plane is:
\begin{equation}
	dk_u\,dk_v
	= \beta^2\,\bigl(\hat{\mathbf{n}}(\mathbf{r})\cdot
	\hat{\mathbf{s}}\bigr)\,d\Omega,
	\label{eq:jacobian_cotangent}
\end{equation}
where $d\Omega=\sin\theta\,d\theta \, d\phi$. Back-facing directions
($\hat{\mathbf{n}}\cdot\hat{\mathbf{s}}\leq 0$) yield a
non-invertible projection and contribute nothing to the radiated
volume. We introduce the notation $[x]_+ = \max(0,x)$ to enforce
this restriction.

To evaluate the total phase-space volume $\Vol_{\mathrm{PS}}$, we
specialise~\eqref{eq:VolPS} to $n=2$, $m=3$. Under the local
flatness assumption $g_{ij} \simeq \delta_{ij}$, the measure on
$\Xi(\mathbf{r})$ reduces to the projected area element on the
forward hemisphere: $d\sigma(\mathbf{k}) = dk_u\,dk_v$ (the
cotangent measure), related to the solid-angle element via the
Jacobian~\eqref{eq:jacobian_cotangent}. The Riemannian measure
reduces to $d\mu_g \simeq dS$. The global integral therefore
becomes:
\begin{equation}
	\Vol_{\mathrm{PS}} = \int_S \left( \int_{\Xi(\mathbf{r})} dk_u\,dk_v \right) dS.
\end{equation}
Substituting the local measure from \eqref{eq:jacobian_cotangent}
and applying the outward-hemisphere restriction $[ \cdot ]_+$,
the integration can be formally extended to the full sphere
$\mathbb{S}^2$ without altering the result:
\begin{equation}
	\Vol_{\mathrm{PS}} = \beta^2 \int_S \int_{\mathbb{S}^2}
	\bigl[\hat{\mathbf{n}}(\mathbf{r}) \cdot \hat{\mathbf{s}}\bigr]_+
	\, d\Omega \, dS.
	\label{eq:VolPS_Geometric_final}
\end{equation}
Geometrically, this integral represents the phase-space footprint of
the field, where the term $[\hat{\mathbf{n}}(\mathbf{r}) \cdot
\hat{\mathbf{s}}]_+$ reduces to $\cos\theta$ for outward-propagating
directions and vanishes otherwise.

The physical connection between the geometric phase-space footprint
and the transport of information is established by examining the
field in the high-frequency limit. In the asymptotic
(geometrical-optics) regime, the WDF converges to the classical
radiance---a correspondence formally developed in Chapter~5
of~\cite{mandel1995optical}.

Specifically, the generalized
radiance $\mathfrak{B}_\nu(\mathbf{r},\hat{\mathbf{s}})$---a
wave-theoretic analogue of spectral radiance derived from field
correlations~\cite[Eq.~(5.7-46)]{mandel1995optical}---converges
pointwise along each ray to the classical radiance
$B_\nu^{(0)}$,\footnote{Classical radiance $B_\nu^{(0)}$ is
	strictly non-negative, in contrast to $\mathfrak{B}_\nu$, which
	may take negative values due to wave interference.} and the
proportionality
\begin{equation}
	\WDFP(\mathbf{r},\mathbf{k})\big|_{|\mathbf{k}|=\beta}
	\propto B_\nu^{(0)}(\mathbf{r}_0,\hat{\mathbf{s}}),
	\qquad \hat{\mathbf{s}}=\mathbf{k}/\beta,
	\label{eq:WDF_ClassB}
\end{equation}
holds locally at each $\mathbf{r}$, where $\mathbf{r}_0$ is the
source point from which the ray along $\hat{\mathbf{s}}$ originates.
In this limit the support of $\WDFP$ is restricted to
$\Xi(\mathbf{r})$, and substituting~\eqref{eq:jacobian_cotangent}
into~\eqref{eq:VolPS} confirms that the phase-space volume is
determined solely by the source geometry, as computed
in~\eqref{eq:VolPS_Geometric_final}. The bridge between the
phase-space description and radiometry is the \textit{\'etendue}
$G_e$~\cite[Chap.~4]{chaves2008introduction}, the radiometric
invariant that quantifies the geometric extent of the
power-transport channel, as shown below.

\begin{figure}[!t]
	\centering
	\includegraphics[width=0.7\linewidth]{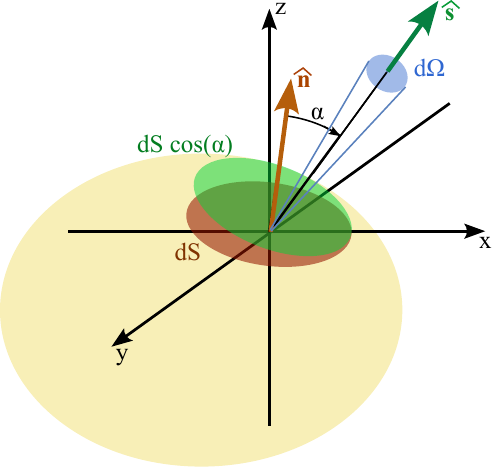}
	\caption{Geometric interpretation of \'etendue.
		A differential surface element $dS$, with normal $\hat{\mathbf{n}}$, radiates
		in the direction $\hat{\mathbf{s}}$, forming an angle $\alpha$.
		The effective projected area is $dS \cos\alpha$, and the
		differential \'etendue is $dG = (dS \cos\alpha)\, d\Omega$.
		The source surface may be closed or open, including the planar case.}
	\label{fig:etend_2}
\end{figure}

The \'etendue is defined by the integral
\begin{equation}
	G_e = \iint dS \cos\alpha \, d\Omega,
\end{equation}
(see Fig.~\ref{fig:etend_2}), and it is conserved under the
geometrical-optics approximation in the absence of diffraction,
absorption, or scattering phenomena.

To connect this quantity to the \NDF\ of the field in the asymptotic
regime, we consider a bounded emitting surface $S$ and the
solid-angle region $\Omega$ into which the source radiates.
We assume observation in the far field, so that every propagation
direction is uniquely represented by a unit vector
$\hat{\mathbf{s}} = \hat{\mathbf{s}}(\theta,\varphi)$.
The corresponding solid-angle element is
$d\Omega = \sin\theta\, d\theta\, d\varphi$.

For each point $\mathbf{r}' \in S$, let $\hat{\mathbf{n}}(\mathbf{r}')$
denote the outward-pointing unit normal. Given a direction
$\hat{\mathbf{s}}$, the scalar product
$\hat{\mathbf{n}}(\mathbf{r}') \cdot \hat{\mathbf{s}}$ provides the
geometric projection factor associated with projecting the surface
element $dS$ onto the plane orthogonal to $\hat{\mathbf{s}}$.
Since only outward-propagating directions contribute to the radiated
field, we use the operator $[x]_+ = \max(0,x)$ introduced
in~\eqref{eq:VolPS_Geometric_final}, so that
$[\hat{\mathbf{n}}(\mathbf{r}') \cdot \hat{\mathbf{s}}]_+$ discards
back-facing contributions.

With these definitions, the geometric \'etendue $G_e$ of the source
is given by
\begin{equation}
	G_e = \int_{\Omega} A_{\mathrm{proj}}(\hat{\mathbf{s}})\, d\Omega,
	\label{eq:Ge_Aproj}
\end{equation}
where $A_{\mathrm{proj}}$ is the projected
area\footnote{Depending on the field of application, the projected
	area is also referred to as the shadow area or cross-sectional
	area. In antenna theory, it is closely related to the concept of
	effective aperture, although the two do not coincide in general,
	as the latter accounts for impedance matching and radiation
	efficiency. In scattering theory, the projected (shadow) area is
	related to scattering and extinction cross sections through the
	optical (extinction) theorem~\cite[Sec.~13.3]{born_wolf_1999};
	in particular, in the high-frequency geometrical-optics limit,
	the extinction cross section of an opaque body converges to twice
	its geometric shadow area.}, which, for convex
surfaces\footnote{For an opaque source surface, this formula can be
	extended to non-convex geometries by supplementing this local
	projection with a global visibility condition. To this end, we
	introduce the visibility function $\chi(\mathbf{r}, \hat{\mathbf{s}})$,
	which is equal to $1$ if the surface point $\mathbf{r}'$ is
	visible from the far field in direction $\hat{\mathbf{s}}$, and
	$0$ otherwise. The projected area is modified as
	$A_{\mathrm{proj}}(\hat{\mathbf{s}}) = \int_S
	\chi(\mathbf{r}, \hat{\mathbf{s}})\,
	[ \hat{\mathbf{n}}(\mathbf{r}) \cdot \hat{\mathbf{s}} ]_+\, dS$,
	where $\chi$ ensures that self-shadowing effects are correctly
	accounted for in arbitrary non-convex geometries.}, is defined as
\begin{equation}
	A_{\mathrm{proj}}(\hat{\mathbf{s}}) = \int_S \bigl[ \hat{\mathbf{n}}(\mathbf{r}) \cdot \hat{\mathbf{s}} \bigr]_+\, dS,
	\label{eq:Aproj_def_main}
\end{equation}
where $dS$ denotes the surface element.

Comparing~\eqref{eq:VolPS_Geometric_final} and~\eqref{eq:Ge_Aproj}:
\begin{equation}
	\Vol_{\mathrm{PS}} = \beta^2\,G_e.
	\label{eq:VolPS_Ge}
\end{equation}

This identity is not a coincidence: the \textit{\'etendue} $G_e$ and
the phase-space volume $\Vol_{\mathrm{PS}}$ are the \emph{same
	object} expressed in two different formal languages. While the
surface element $dS$ represents the Riemannian measure on the tangent
space $T_{\mathbf{r}}M$, the product
$\cos\theta\,d\Omega\,dS = \beta^{-2}d^2q\,d^2k_{\parallel}
= \beta^{-2}\varpi^2/2$ is precisely the symplectic area form on the
four-dimensional phase space $T^*M$ (for $n=2$).

Consequently, the identity $G_e = \beta^{-2}\Vol_{\mathrm{PS}}$
follows directly from the fundamental geometry of the cotangent
bundle. Within this framework, the conservation of \textit{\'etendue}
during propagation is an immediate consequence of Liouville's theorem
applied to the invariance of the symplectic form
$\varpi^2$~\cite{chaves2008introduction, markvart2007thermodynamics}.
This geometric invariance also corresponds to the conservation of
optical entropy~\cite{markvart2007thermodynamics} in lossless systems.

Dividing by the Heisenberg cell $(2\pi)^2$:
\begin{equation}
	\text{NDF} = \frac{G_e}{\lambda^2}.
	\label{eq:NDF_final}
\end{equation}
It is worth noting that a completely equivalent expression was previously derived by Gustafsson in \cite{gustafsson2025degrees} formulating the NDF in terms of the total shadow area. The explicit connection between the shadow area and the radiometric étendue will be rigorously formalized below, proving that they represent different manifestations of the same underlying symplectic invariant. 

As an example, for a sphere of radius $a$: $A_{\mathrm{proj}}=\pi a^2$, $G_e=4\pi^2 a^2$, and $\text{NDF}=(\beta a)^2$. For a planar area $A$, approaching it as the limit of a closed surface yields $G_e=2\pi A$ ($\text{NDF}=2\pi A/\lambda^2$) due to bidirectional radiation. Conversely, restricting propagation to a single half-space halves the spectral support $\Xi(r)$, giving $G_e=\pi A$ and $\text{NDF}=\pi A/\lambda^2$, consistent with directional aperture formulations \cite{gustafsson2025degrees}.

Before concluding this section, three observations are in order,
each highlighting a different facet of the same underlying symplectic
structure.

First, the \'etendue is a \emph{purely geometrical} property of the
source--observer configuration. The coherence properties of the source
do not alter $G_e$; they determine how efficiently the available
phase-space volume is filled. For Lambertian
sources,\footnote{The Lambertian assumption is justified by the
	Maximum Entropy Principle: for given radiated power, a Lambertian
	source maximizes the occupied phase-space volume.}
$\Vol_{\mathrm{PS}}$ is fully populated and \eqref{eq:NDF_final}
gives the exact \NDF. This explains why the \NDF\ results for
deterministic and stochastic sources align, and why the
identification of the \NDF\ in the asymptotic regime reduces to a
purely geometrical problem: the symplectic volume
$\Vol_{\mathrm{PS}}=\beta^2 G_e$ is fixed by the geometry of the
source, independently of the source statistics.

Second, the \textit{total shadow area}, used in~\cite{gustafsson2025degrees} to evaluate the NDF under asymptotic conditions, is obtained by integrating the projected area over all radiated directions,
\begin{equation}
	A_{\mathrm{shadow}} = \int_{\Omega}
	A_{\mathrm{proj}}(\hat{\mathbf{s}})\,d\Omega,
	\label{eq:Ashadow}
\end{equation}
which is precisely the defining integral of the geometric \'etendue $G_e$ in~\eqref{eq:Aproj_def_main}; the two quantities therefore coincide.

From the symplectic perspective this is not a coincidence: the
shadow area is the spatial projection of the symplectic volume
$\Vol_{\mathrm{PS}}=\beta^2 G_e$, and its invariance under
propagation is a direct consequence of the preservation of $\varpi$
under Hamiltonian flow. The three quantities---total shadow area,
\'etendue, and phase-space volume---are therefore different faces of
the same symplectic invariant: geometric, radiometric, and
phase-space, respectively.

Third, the factorization $\Vol_{\mathrm{PS}} =
A_{\mathrm{eff}}\cdot|\Xi|$, where
\begin{equation}
	A_{\mathrm{eff}} = \frac{G_e}{\Omega'},
	\qquad
	\Omega' = \frac{|\Xi|}{\beta^2} = \pi,
	\label{eq:Aeff_def}
\end{equation}
and $|\Xi|=\pi\beta^2$ is the spectral support area in the cotangent plane, gives:
\begin{equation}
	\NDF = \frac{A_{\mathrm{eff}}\cdot|\Xi|}{(2\pi)^2}.
	\label{eq:NDF_SBP_final}
\end{equation}
Here $A_{\mathrm{eff}}$ has a precise geometric interpretation:
it is the area of the source surface as seen through the
symplectic measure of the cotangent bundle, i.e., the
projected area weighted by $\cos\theta\,d\Omega/\pi$ rather
than by the uniform solid-angle measure $d\Omega/2\pi$. It
lives naturally on the tangent bundle $TM$---it is a measure
of spatial extent---while $|\Xi|$ lives on the cotangent fiber
$T^*_{\mathbf{r}}M$---it is a measure of spectral support.
Together they form a Fourier-conjugate pair on the cotangent
bundle $T^*M$, in exact analogy with the spatial
extent--bandwidth product of Landau's
theory~\cite{landau1961prolate, slepian1962prolateIII}, which
applies to planar geometries in Euclidean space.

The extension of this space--bandwidth factorization to curved
surfaces via the cotangent bundle framework, as presented here,
does not appear to have been stated explicitly in the
literature in this form, to the authors' knowledge, although
related results for the shadow area appear
in~\cite{gustafsson2025degrees} and extensions of the Landau
argument to general domains in Euclidean space are given
in~\cite{franceschetti2015landau}. The key point of the
present formulation is that the curvature of $M$ drops out
exactly---through the metric cancellation
$\det(g_{ij})\cdot\det(g^{ij})=1$ established
in~\eqref{eq:Liouville}---so that the Landau counting argument
applies intrinsically on any smooth manifold $M$, without
requiring a planar approximation. This provides the
final link in the symplectic chain: the Fourier conjugacy of
$A_{\mathrm{eff}}$ and $|\Xi|$ is precisely the content of
the symplectic pairing $\varpi$ between positions and
wavevectors, and the elementary cell $(2\pi)^2$ is the
symplectic area of one mode.

\section{Spatial Information as the Physical Footprint of Hamiltonian Structure}
\label{sec:Line_source}

This final section provides the physical unification of deterministic and stochastic frameworks by demonstrating how the abstract symplectic geometry of phase space leaves a tangible ``footprint'' in physical space: \textit{the Spatial Information Flows (SIFs)}. We show that what is perceived as the spatial distribution of information is the physical manifestation of the underlying Hamiltonian structure of the field. While the general theory of Sections~\ref{app:functional_models}--\ref{sec:EIT_modelB} applies to arbitrary geometries, we illustrate this principle here through the canonical case of a spatially incoherent linear source of length $\ell=2a$.

\subsection{CSD Analysis and the Emergence of \SIFs}

The CSD for a spatially uncorrelated, homogeneous line source
with $S_J=1$ on $-a\le y'\le a$ is:
\begin{equation}
	W(\mathbf{r}_1,\mathbf{r}_2)
	= \frac{(\bar\omega\mu_0)^2}{16\pi^2}
	\int_{-a}^{a}
	\frac{e^{j\beta[R_2(y')-R_1(y')]}}{R_1(y')\,R_2(y')}\,dy',
	\label{eq:W1}
\end{equation}
where $R_n(y')=\sqrt{z_n^2+(y_n-y')^2}$.\footnote{This
	section works with the electric-field cross-spectral density
	$W(\mathbf{r}_1,\mathbf{r}_2)$ rather than the power
	cross-spectral density $W_p$ of Section~\ref{sec:EIT_modelB}.
	For the spatially incoherent source considered here
	($S_J=\mathrm{const}$, $L_c\to 0$), the two are related by
	$W_p \approx W/2\zeta_0$ whenever $R\gg\lambda$, because
	the cross terms between different plane-wave components
	cancel exactly by statistical averaging. This condition is
	satisfied throughout the radiation zone considered here.
	All structural results---the sinc coherence, the \SIF\
	geometry, and the connection with Bucci's sampling---are
	therefore independent of this choice.}
The phase of the integrand is the \emph{difference of classical
	actions}:
\begin{equation}
	\phi(y')
	= \beta[R_2(y')-R_1(y')]
	= \mathcal{S}(\mathbf{r}_2,y') - \mathcal{S}(\mathbf{r}_1,y'),
	\label{eq:phase_action}
\end{equation}
where $\mathcal{S}(\mathbf{r}_n,y')=\beta R_n(y')$ is the classical
action along the ray from source point $y'$ to observation point
$\mathbf{r}_n$. The CSD is therefore the coherent superposition of
action differences, integrating over all source points.

Using the linear phase approximation\footnote{The linear interpolation error for each $R_n(y')$
		individually is $\mathcal{O}(a^2/z_n)$, which can be large
		in the Fresnel zone. However, in the \emph{difference}
		$R_2(y')-R_1(y')$ the dominant nonlinear terms nearly
		cancel when the two observation points are close
		($|\delta\mathbf{r}|\ll R$), leaving a residual of order
		$\mathcal{O}(a^2|\delta z|/z^2 + a^3|\delta y|/z^3)$.
		This is negligible whenever $|\delta\mathbf{r}|\ll z$,
		which is precisely the regime in which the two-point
		coherence function is evaluated.}:
\begin{equation}
	R_n(y') \approx \frac{R_n^++R_n^-}{2}
	+ y'\frac{R_n^+-R_n^-}{2a},
	\label{eq:linear_phase_approx}
\end{equation}
where $R_n^\pm=\sqrt{z_n^2+(y_n\mp a)^2}$ are the distances from the
two source endpoints to $\mathbf{r}_n$, one obtains:
\begin{align}
	W(\mathbf{r}_1,\mathbf{r}_2)
	&\approx \frac{(\bar\omega\mu_0)^2}{16\pi^2}
	\frac{2a}{z_1z_2}\,e^{j\beta\Phi_{12}}
	\nonumber\\
	&\quad\times
	\sinc\!\left(\beta
	\frac{(R_2^+-R_2^-)-(R_1^+-R_1^-)}{2}\right),
	\label{eq:W_sinc}\\
	\Phi_{12}
	&= \frac{R_2^++R_2^-}{2} - \frac{R_1^++R_1^-}{2},
	\label{eq:Phi12_def}
\end{align}
where $\sinc(x)=\sin x/x$. The spectral degree of coherence is:
\begin{equation}
	|\mu(\br_1;\br_2)|
	\approx \sinc\!\left(\beta
	\frac{(R_2^+-R_2^-)-(R_1^+-R_1^-)}{2}\right).
	\label{eq:mu_sinc}
\end{equation}

The argument of the sinc depends only on the difference
of the action difference $\Delta\mathcal{S}$ between the
two observation points:
\begin{equation}
	\Delta\mathcal{S}(\mathbf{r}_n)
	= \beta(R_n^+ - R_n^-),
	\label{eq:DeltaS}
\end{equation}
where $\Delta\mathcal{S}(\mathbf{r}_n)$ is the difference
between the classical actions accumulated along the two
endpoint rays reaching $\mathbf{r}_n$---one from
$\mathbf{r}_+=(0,+a)$ and one from $\mathbf{r}_-=(0,-a)$.
Therefore $|\mu(\mathbf{r}_1;\mathbf{r}_2)|$ is a function
only of $\Delta\mathcal{S}(\mathbf{r}_2) -
\Delta\mathcal{S}(\mathbf{r}_1)$. 
In particular,
$|\mu(\mathbf{r}_1;\mathbf{r}_2)| \approx 1$ (maximum coherence, within the sinc approximation)
when $\Delta\mathcal{S}(\mathbf{r}_2)=
\Delta\mathcal{S}(\mathbf{r}_1)$, i.e., when the two
points lie on the same \SIF. 
The sinc vanishes when
$|\Delta\mathcal{S}(\mathbf{r}_2)-
\Delta\mathcal{S}(\mathbf{r}_1)| = 2m\pi$ for any
non-zero integer $m$, corresponding to a path-difference
change of $m\lambda$ in the physical distances $R^\pm$.
The coherence length is identified with the
first zero ($m=1$): two points for which the action
difference $\Delta\mathcal{S}$ differs by $2\pi$
(equivalently, for which $(R^+-R^-)$ differs by $\lambda$)
are uncorrelated, and points
separated by more than $\lambda$ can be considered
statistically independent. On the source surface, where
$\Delta\mathcal{S}(y_0)=-2\beta y_0$ varies linearly with
position, this coherence length maps to a source spacing
of $\lambda/2$---in exact agreement with Bucci's Nyquist
sampling criterion.
The curves  $\Delta\mathcal{S}(\mathbf{r}) = \beta(R^+ - R^-) = \text{const}$---along which the coherence decays most slowly---are hyperbolas with foci at the source endpoints. These loci of slowest decorrelation, and consequently of high mutual information (i.e.\ redundancy) between points belonging to the same locus, are the \textit{Spatial Information Flows} (\SIFs) introduced in~\cite{migliore2023classical,migliore2022information}.
The \SIFs\ are the asymptotic level sets of the \textit{action difference} $\Delta\mathcal{S}=\beta(R^+-R^-)$, built from the actions of two distinct rays---one from each source endpoint---reaching the observation point. They are \textit{not} ray trajectories: the tangent to a level set of $\Delta\mathcal{S}$ at $\mathbf{r}$ is orthogonal to
\begin{equation}
	\nabla\Delta\mathcal{S}(\mathbf{r})
	= \beta\!\left(
	\frac{\mathbf{r}-\mathbf{r}_+}{|\mathbf{r}-\mathbf{r}_+|}
	-\frac{\mathbf{r}-\mathbf{r}_-}{|\mathbf{r}-\mathbf{r}_-|}
	\right)
	= \beta(\hat{\mathbf{s}}^+-\hat{\mathbf{s}}^-),
	\label{eq:gradDeltaS}
\end{equation}
the difference of the unit vectors pointing from the two source endpoints to $\mathbf{r}$. This direction is in general not aligned with the local ray direction $\mathbf{k}/|\mathbf{k}|$, which represents the dominant propagation direction at $\mathbf{r}$.

Using an analogy, \textit{rays} are like the trajectories of individual water molecules flowing along a river, while SIFs represent the \textit{coherence structure} of the river itself---defined by the interference between the two banks (the source endpoints). The river flows along the rays, but information is organized across the SIFs. The two structures coincide only asymptotically in the far field, where the rays from the two endpoints become nearly parallel ($\hat{\mathbf{s}}^+ \approx \hat{\mathbf{s}}^-$). In this limit, the transverse component of $\nabla\Delta\mathcal{S} = \beta(\hat{\mathbf{s}}^+ -
\hat{\mathbf{s}}^-) \to \mathbf{0}$, and the level sets
of $\Delta\mathcal{S}$ elongate in the common radial
direction, causing the SIFs to align with the rays.

The action difference $\Delta\mathcal{S}$ is the geometric 
counterpart of the Wigner separation variable $\Delta\mathbf{r}$: 
just as the WDF depends on $\Delta\mathbf{r}$ paired with 
$\mathbf{k}_{\parallel}$ via the canonical cotangent pairing 
$k_{\parallel,i} \, dq^i$, the CSD depends on $\Delta\mathcal{S}$, 
	whose difference between the two observation points, 
	$\Delta\mathcal{S}(\mathbf{r}_2)-\Delta\mathcal{S}(\mathbf{r}_1)$, 
	is the natural two-point separation.

The subset $\mathbf{Q}_{\mathrm{SIF}}$ of \SIFs\ separated by
$2\pi$ in action difference carries statistically independent
information: two points on adjacent \SIFs\ have $|\mu|=0$ in the
sinc approximation. This subset defines the independent information
channels of the radiation field.

\subsection{The \SIFs\ as Cotangent Curves: Connection with
	Deterministic Optimal Sampling Theory}
\label{subsec:cotangent}

\begin{figure}
	\centering
	\includegraphics[width=0.8\linewidth]{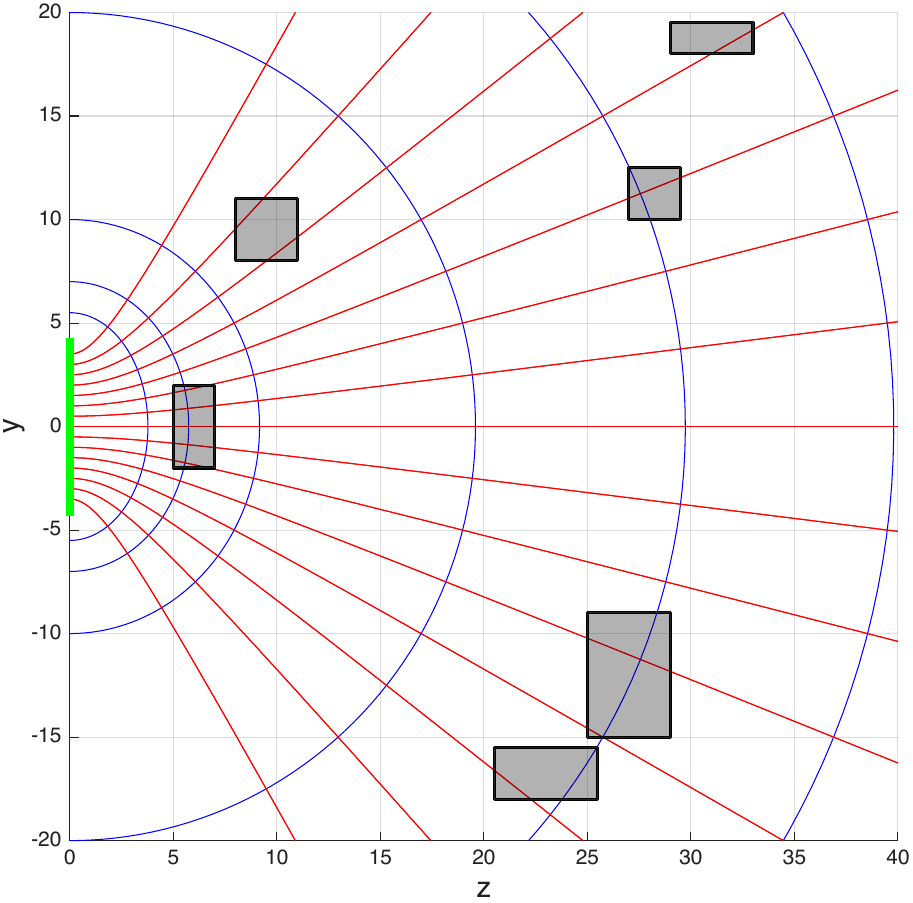}
	\caption{The $\xi$ hyperbolas (red curves) for a linear source of
		length $\ell=8\lambda$ (green line). Gray rectangles are receiving
		domains. The $\xi$ curves are simultaneously the optimal sampling
		curves of Bucci et al.~\cite{bucci1998representation}, the
		\SIFs\ of the field, and the iso-cotangent curves of the
		radiation channel.}
	\label{fig:urbanov2a}
\end{figure}

The \SIF\ structure, derived above from the properties of the CSD, identifies a 
fundamental connection with the optimal sampling theory developed by Bucci, 
Franceschetti, and co-workers~\cite{bucci1987spatial, bucci1998representation}. The warping coordinate 
$\xi$ of Bucci is defined as (eqs.~(17)--(18) 
of~\cite{bucci1998representation}):
\begin{equation}
	\xi = \frac{\beta}{\mathcal{B}} \left[
	\frac{R_1 - R_2}{2} + \frac{s'_1 + s'_2}{2}
	\right] + \text{const},
	\label{eq:xi_Bucci}
\end{equation}
where $R_{1,2}$ are distances from the two \textit{tangency points} 
of the convex source surface $S$ to the observation point, 
$s'_{1,2}$ are their arclength coordinates on $S$, and $\mathcal{B}$ 
denotes the effective bandwidth of the reduced field representation. 
For the linear source, the tangency points are the two \textit{fixed}
endpoints $y' = \pm a$ for every $\mathbf{r}$, so $R_{1,2} = R^\pm$
and $\frac{s'_1 + s'_2}{2}$ is constant; hence $\xi$ is an affine
function of $\Delta\mathcal{S}(\mathbf{r}) = \beta(R^+ - R^-)$.
The constant-$\xi$ curves of Bucci are therefore identical 
to the constant-$\Delta\mathcal{S}$ curves---i.e., the 
SIFs (Fig.~\ref{fig:urbanov2a}).
This identification is not a coincidence: both constructions single
out the same geometric object---the level sets of the action difference
between the two boundary rays of the source---from two different
physical perspectives. Bucci arrives at $\xi$ by requiring the local
sampling bandwidth to be constant along the observation curve
(an optimality condition in approximation theory). The SIFs arise from requiring maximum mutual coherence between field points (a condition
in information theory). The fact that both lead to the same curves
is a manifestation of the deep connection between field representation
and information transport.

The key geometric insight is that, for the 2D geometry considered
here, $\xi$ is the natural cotangent coordinate along the observation
curve: the action difference $\beta(R^+-R^-)$ represents the phase
difference between the two endpoint rays reaching $\mathbf{r}$,
effectively measuring the ``spectral distance'' in the cotangent
space $T^*_{\mathbf{r}}M$. Uniform sampling at intervals
$\Delta\xi=\pi/\mathcal{B}$ corresponds to uniform spacing $\lambda/2$
on the source, consistent with the canonical phase-space cell
$(2\pi)$. In the physical observation plane the $\xi$-grid is
non-uniform: the SIFs are hyperbolas whose spacing varies with
position (Fig.~\ref{fig:urbanov2a}).

\begin{figure}
	\centering
	\includegraphics[width=0.8\linewidth]{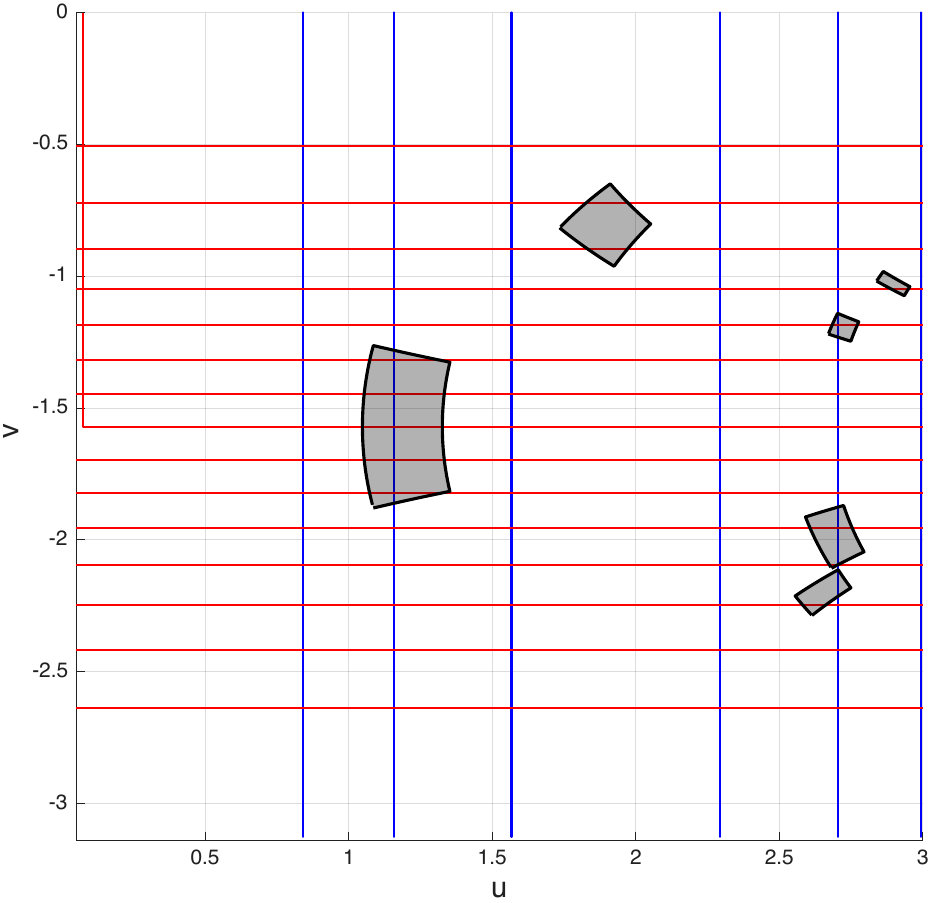}
	\caption{Warped coordinate space: the source perceives its
		surroundings through $w(\zeta)=\mathrm{arccosh}(-i\zeta/a)$,
		$\zeta=y+iz$. This is the ``electrically flat'' space of
		Section~\ref{sec:EIT_modelB}: the coordinate system in which
		$g_{ij}\simeq\delta_{ij}$ everywhere, so that the symplectic
		cell $2\pi$ has the same area at every point. The \SIFs\ become
		straight lines; their uniform spacing reflects the metric
		independence of the elementary phase-space cell.}
	\label{fig:urbanov2b}
\end{figure}

This explains why Bucci's sampling is optimal for the 2D geometry
	considered here: the $\xi$ samples lie on the \SIFs, the natural
information-carrying structures of the radiation channel. Adjacent
samples on different \SIFs\ carry statistically independent
information, while samples on the same \SIF\ are redundant. 
The warped coordinate system $w(\zeta)=\mathrm{arccosh}(-i\zeta/a)$,
$\zeta=y+iz$ (Fig.~\ref{fig:urbanov2b}), linearises the \SIFs\
because each \SIF\ (a level set of $\xi$) maps to a line
	$\mathrm{Im}(w)=\mathrm{const}$. This is the coordinate chart in
which $g_{ij}\simeq\delta_{ij}$ everywhere---the ``electrically flat''
space of Section~\ref{sec:EIT_modelB}---and the uniform spacing of the
\SIFs\ reflects the metric independence of the elementary phase-space
cell $2\pi$ (Eq.~\eqref{eq:Liouville}).%
\footnote{Currently, a global optimal warping transformation is
	available only along observation
	curves~\cite{bucci1997electromagnetic}. Its extension to surface
	observation domains remains an active area of
	research~\cite{solimene2019sampling}.}

A deeper geometric interpretation of field coherence arises from the
cotangent bundle $T^*M$ of the observation manifold. In the
geometrical-optics (high-frequency) limit, the wavevectors of the
rays reaching $\mathbf{r}$ from the two source endpoints
$\mathbf{r}^\pm = (0,\pm a)$ are $\mathbf{k}^\pm =
\beta\hat{\mathbf{s}}^\pm$. Their projections onto the cotangent
fiber $T^*_{\mathbf{r}}M$ along the local unit tangent
$\hat{\mathbf{t}} \in T_{\mathbf{r}}M$ are:
\begin{equation}
	k_\parallel^\pm
	= \mathbf{k}^\pm\cdot\hat{\mathbf{t}}
	= \beta\,\hat{\mathbf{s}}^\pm\cdot\hat{\mathbf{t}}
	= \beta\cos\theta^\pm
	= \beta\frac{dR^\pm}{ds},
	\label{eq:projections}
\end{equation}
where $s$ is the arclength along $M$, $\theta^\pm$ are the angles
between the endpoint rays $\hat{\mathbf{s}}^\pm$ and
$\hat{\mathbf{t}}$, and the last equality follows from the geometric
identity $\frac{dR^\pm}{ds} = \hat{\mathbf{s}}^\pm\cdot\hat{\mathbf{t}}
= \cos\theta^\pm$ (the rate of change of the distance to a fixed
point is the cosine of the angle between the direction to that point
and the direction of motion). The two projected values
$k_\parallel^\pm$ are the local spatial frequencies associated with
the two endpoint rays sensed along the manifold; their separation is
the quantity of physical interest.
The width of the projected two-ray fan in the cotangent fiber defines
the \textit{local spatial bandwidth}:
\begin{equation}
	W_k = |k_\parallel^+ - k_\parallel^-|
	= \beta|\cos\theta^+ - \cos\theta^-|.
\end{equation}
Physically, $W_k$ is the local density of degrees of freedom per unit
arclength of $M$: the number of independent spatial samples
intercepted along a portion of $M$ is $\frac{1}{2\pi}\int W_k\,ds$.
This mirrors the time--bandwidth product of classical signal theory,
with spatial displacement $ds$ playing the role of observation time
$dt$; Table~\ref{tab:analogy} makes the parallel explicit.

\begin{table}[h]
	\centering
	\caption{Time vs. Spatial Information Analogy}
	\label{tab:analogy}
	\begin{tabular}{lcc}
		\hline
		\textbf{Feature} & \textbf{Time Domain} & \textbf{Spatial Domain} \\ \hline
		Variable & Time ($t$) & Arclength ($s$) \\
		Conjugate & Ang. Freq. ($\omega$) & Wavevector ($k_\parallel$) \\
		Spectrum & Bandwidth ($B_\omega$) & Local Fan ($W_k$) \\
		Uncert. Cell & $\Delta \omega \Delta t \geq 2\pi$ & $\Delta k_\parallel \Delta s \geq 2\pi$ \\ \hline
	\end{tabular}
\end{table}

\textit{Moving along a SIF} ($W_k = 0$). The SIFs are the curves along
which $\Delta\mathcal{S} = \beta(R^+ - R^-) = \text{const}$, so
$d\Delta\mathcal{S}/ds = 0$ and the path difference between the two
endpoint rays remains fixed. The two cotangent projections then
coincide, $k_\parallel^+ = k_\parallel^-$, and the fan collapses to a
single point: the observer stays in phase synchrony with the
interference pattern set by the two source endpoints. This is the
condition of maximum coherence ($|\mu|\to 1$ as the sinc argument
	vanishes) and maximum redundancy: no new spatial frequency is
intercepted, and no new degree of freedom is accumulated.

\textit{Moving across SIFs} (orthogonal to a SIF, $W_k > 0$).
Displacing $\mathbf{r}$ across the SIF family changes
$\Delta\mathcal{S}$ and shifts the two projections $k_\parallel^\pm$ by
different amounts: the endpoint rays fall out of synchronization and
the field becomes partially decorrelated. When the accumulated phase
difference reaches $2\pi$---i.e., $|\Delta\mathcal{S}(\mathbf{r}_2) -
\Delta\mathcal{S}(\mathbf{r}_1)| = 2\pi$ (equivalently, the path
difference $(R^+-R^-)$ changes by $\lambda$)---the observation domain
has crossed one full Heisenberg cell in the $(s, k_\parallel)$ phase
plane, intercepting one new independent degree of freedom. This is
exactly when the sinc in~\eqref{eq:mu_sinc} first vanishes.

The accumulated decorrelation between two points $\mathbf{r}_1$ and
$\mathbf{r}_2$ connected by a path crossing the SIF family is
therefore measured by the integrated local bandwidth along that path.
Projecting $\nabla\Delta\mathcal{S} = \beta(\hat{\mathbf{s}}^+ -
\hat{\mathbf{s}}^-)$ (see Eq.~\eqref{eq:gradDeltaS}) onto the
displacement $d\mathbf{r}$, one obtains:
\begin{equation}
	\frac{\Delta\mathcal{S}(\mathbf{r}_2) - \Delta\mathcal{S}(\mathbf{r}_1)}{2}
	= \frac{1}{2}\int_{\mathbf{r}_1}^{\mathbf{r}_2}
	\nabla\Delta\mathcal{S}\cdot d\mathbf{r},
	\label{eq:sinc_bandwidth}
\end{equation}
where $|\nabla\Delta\mathcal{S}\cdot d\mathbf{r}| \approx
	|\nabla\Delta\mathcal{S}|\,ds = W_k\,ds$ when the path is oriented
predominantly orthogonal to the SIFs (i.e., along
$\nabla\Delta\mathcal{S}$). This is exactly the argument of the sinc
in~\eqref{eq:mu_sinc}, confirming that the NDF intercepted by an
observation domain is the integral of the local bandwidth $W_k$ along
the domain, divided by $2\pi$---the spatial analogue of Shannon's
time-bandwidth product, here formulated intrinsically on the cotangent
bundle $T^*M$ of the observation manifold.

\subsection{Numerical Verification and Phase-Space Invariance of the
	\SIFs}

The above discussion is based on an approximate correlation. To
identify the actual zone of high mutual information around
$\mathbf{r}_1$, we numerically estimate the CSD and mutual
information.
\begin{figure}
	\centering
	\includegraphics[width=1.0\linewidth]{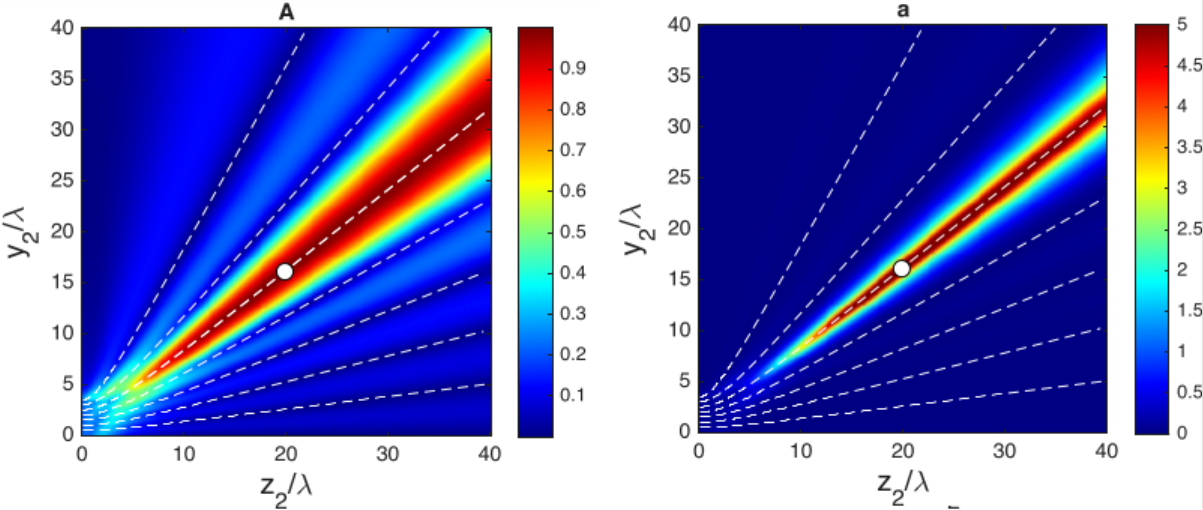}
	\caption{Spectral degree of coherence $|\mu(\br_1;\br_2)|$ (left)
		and mutual information $I_{MI}(\mathbf{r}_1,\mathbf{r}_2)$
		(right) for $\ell=8\lambda$. $\mathbf{r}_1$ in the far field
		(white circle). Dashed curves: $\mathbf{Q}_\xi$ hyperbolas,
		coinciding with the \SIFs.}
	\label{fig:fignspctral1}
\end{figure}
\begin{figure}
	\centering
	\includegraphics[width=1.0\linewidth]{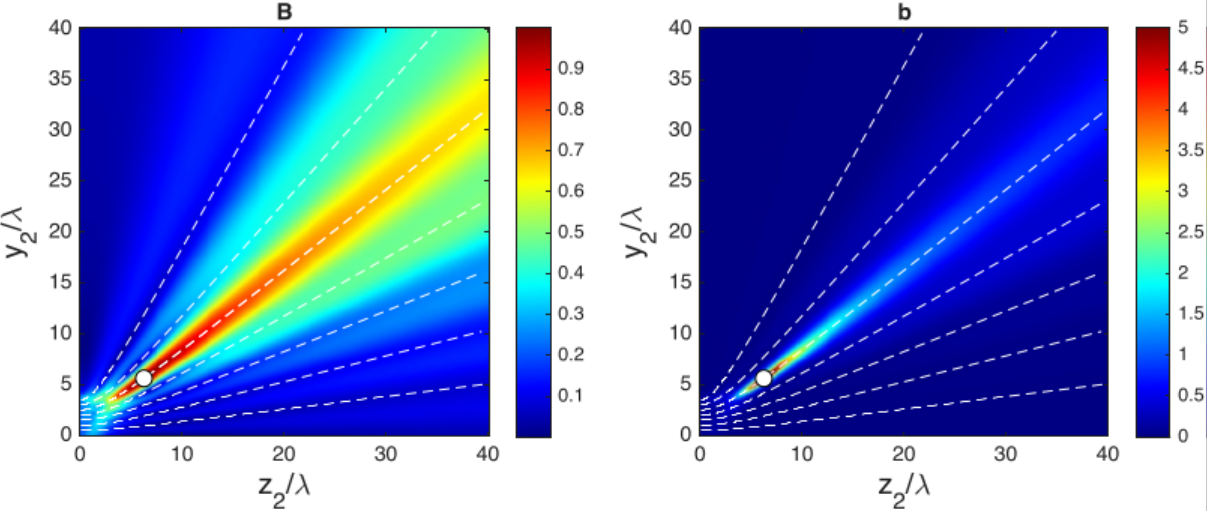}
	\caption{Same as Fig.~\ref{fig:fignspctral1} with $\mathbf{r}_1$
		in the Fresnel zone. The \SIFs\ and $\xi$ hyperbolas remain
		aligned throughout, confirming the invariance of the cotangent
		structure under propagation.}
	\label{fig:fignspctral2}
\end{figure}
Figure~\ref{fig:fignspctral1} shows $|\mu(\mathbf{r}_1;\mathbf{r}_2)|$
and $I_{MI}(\mathbf{r}_1,\mathbf{r}_2)$ with $\mathbf{Q}_\xi$
hyperbolas overlaid. The results confirm the emergence of a
high-mutual-information path---a \SIF---along the hyperbola
passing through $\mathbf{r}_1$, and the near-vanishing of mutual
information between adjacent \SIFs.
In the asymptotic regime, the \SIFs\ form a family of information flux
tubes whose total phase-space footprint is $\beta\,G_{e,\text{1D}}$;
each tube occupies one Heisenberg cell $2\pi$, so their count
equals the \NDF~\cite{migliore2006role}. To verify \NDF\ invariance
beyond the far field, we repeat the analysis with $\mathbf{r}_1$ in the
Fresnel zone (Fig.~\ref{fig:fignspctral2}). The statistical quantities
and the $\xi$ hyperbolas follow the same spatial pattern throughout,
pointing to the deeper Hamiltonian structure discussed next.

\begin{figure}
	\centering
	\includegraphics[width=0.9\linewidth]{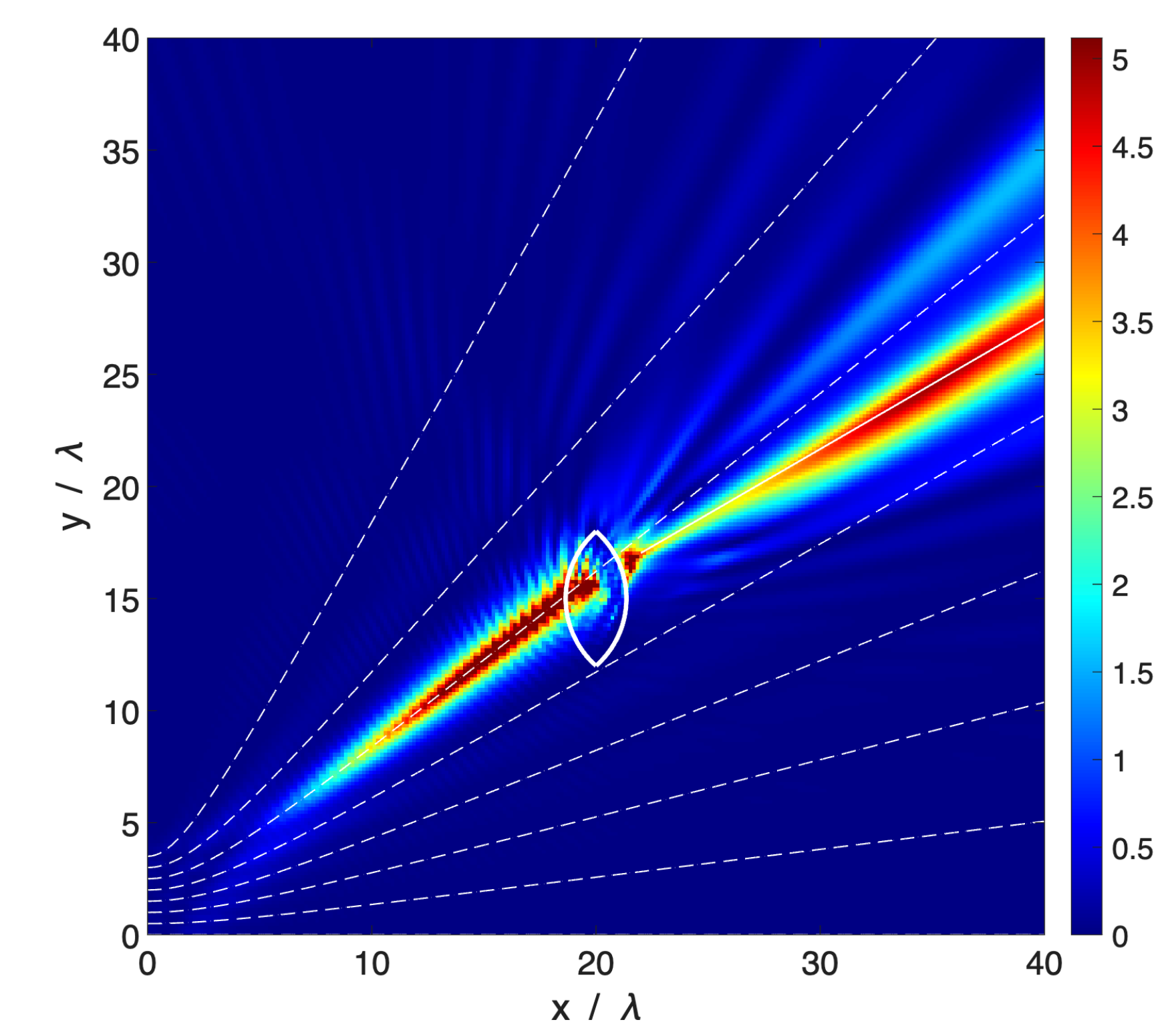}
	\caption{SIF manipulation using a dielectric lens ($\epsilon_r=1.7$).
		Mutual information in false color; refracted optical ray as solid
		line.}
	\label{fig:figlentev3}
\end{figure}

The geometric structure of \SIFs\ is rooted in two complementary
principles from the Hamiltonian framework~\cite{zworski2012semiclassical,
	frankel2004geometry}.
The \textit{shape} of \SIFs\ is determined by the difference of the two endpoint eikonals, each a solution of the
Hamilton--Jacobi equation $\mathscr{H}(\mathbf{r},\mathbf{k})=|\mathbf{k}|^2-\beta^2 =0$ (see SM, Section~II).  \SIFs\
are \textit{not} ray trajectories: they are level sets of the action
difference $\Delta\mathcal{S}=\beta(R^+-R^-)$, built from the actions
of two distinct rays. \SIFs\ bend near the source, where
$\Delta\mathcal{S}$ varies rapidly, and become parallel to the rays
only in the far field---consistently with the river analogy introduced
in Section~\ref{sec:Line_source}: the river flows along the rays, but
the information is organised across the \SIFs.
The \textit{invariance} of \SIFs\ follows from Liouville's theorem:
the phase-space volume element $d\mathbf{r}\,d\mathbf{k}$ is
conserved under the Hamiltonian flow, so the phase-space measure of
the ray bundle connecting consecutive \SIFs\ is preserved during
propagation, and the \NDF\ count between them remains invariant from
the Fresnel zone to the far field. This microscopic conservation
underlies the macroscopic invariance of $G_e$.
Since the \SIFs\ are governed by the same Hamiltonian controlling
power transport, they can be manipulated by lenses, reflectors, and
scattering structures. Fig.~\ref{fig:figlentev3} illustrates this with
a dielectric lens ($\epsilon_r=1.7$): the lens modifies the accumulated
action and reshapes the \SIFs, while Liouville's and Gromov's theorems
guarantee that their total count---the \NDF---is preserved. The lens
redistributes the information flows without creating or destroying
independent channels. Spatial information can therefore be engineered
at the \textit{Deep Physical Layer}~\cite{migliore2021world}, whose
natural geometry is symplectic, with the \SIFs\ as invariant structures
and the \'etendue as capacity measure.

In summary: the Hamiltonian action difference determines \textit{where}
the \SIFs\ are; Liouville's theorem guarantees \textit{how many} reach
the observer; and the cotangent coordinate $\xi$ is the natural
variable in which both the CSD and Bucci's sampling are expressed.
In more complex environments---inhomogeneous media or scattering---these
principles remain valid, with the \SIFs\ mirroring the full propagation
history.
The Hamiltonian structure of \SIFs\ has direct consequences for
channel capacity. Each \SIF\ carries almost independent information
with respect to the others. In the asymptotic regime, where the
significant singular values of the radiation operator are approximately
equal before the exponential cutoff, the channel reduces to $\NDF$
parallel Gaussian sub-channels of equal gain,
giving~\cite{migliore2006role}:
\begin{equation}
	\mathcal{C} \approx \NDF \cdot \log_2(1+\mathrm{SNR}),
\end{equation}
where SNR is the per-channel signal-to-noise ratio. The approximation
reflects two idealizations: the equal-gain assumption on the singular
values, and the residual mutual information between adjacent \SIFs\
(never rigorously zero, as discussed below).
However, channel capacity alone does not fully capture the nature of
\SIFs. Unlike the discrete, orthogonal channels of the
Hilbert--Schmidt or Karhunen--Lo\`eve expansions, \SIFs\ are
continuous information flows that permeate the entire propagation
volume. In practice, a \SIF\ is identified by fixing a threshold on
$|\mu|$, and this threshold-dependence is not merely a practical
limitation but a fundamental consequence of the analyticity of the
radiated field: since $W(\mathbf{r}_1,\mathbf{r}_2)$ satisfies the
Helmholtz equation in each variable separately in the source-free
region, it is analytic in both arguments~\cite{colton1998inverse}, and
$|\mu|$ cannot vanish on any set of positive measure without vanishing
identically. The mutual information between points on different \SIFs\
is therefore never rigorously zero, and the boundaries between
independent channels are inherently diffuse rather than sharp.
This spatial structure of mutual information goes beyond the standard
Shannon capacity framework, which characterises the aggregate channel
between transmitter and receiver but not \textit{where} in the
propagation volume independent spatial channels are available. The
\SIF\ framework could fill this gap through an \textit{information
	coverage map} complementary to the classical power coverage maps used
in wireless network planning---of particular relevance for
spatial-diversity systems such as massive MIMO, distributed antennas,
and cell-free architectures. Whether this potential can be realised in
practical deployments, with complex propagation, multipath, and losses,
requires further research beyond the scope of this paper.

\section{Conclusions}
\label{sec:Conclusions}

This paper has unified deterministic and stochastic Electromagnetic
Information Theory (EIT) by identifying the common structural
skeleton underlying both formulations. Our analysis demonstrates
that the long-observed convergence of these two models---yielding
identical eigenvalues, basis functions, and \NDF---is not a
mathematical coincidence but a structural necessity dictated by the
symplectic geometry of the source--observer phase space.

The core contribution is the identification of a single geometric
entity appearing under different guises in distinct disciplines:
the \'etendue of radiometry, the phase-space volume of Hamiltonian
mechanics, the total shadow area,  and the \NDF\ of the electromagnetic theory are the same
symplectic invariant.

Physically, this symplectic structure manifests through the
formation of \textit{Spatial Information Flows} (\SIFs)---level-set
curves along which the local bandwidth in the cotangent fiber
collapses to zero, rendering nearby field values redundant. These
information tracks correspond to level sets of the classical action
difference; for convex sources they asymptotically coincide with the optimal
sampling curves of Bucci et al. The cross-spectral density depends
solely on the difference of phase-space coordinates along these
tracks---analogous to the Wigner separation variable---providing
the physical signature of the underlying geometric skeleton.

The symplectic framework provides two fundamental constraints on the
radiation channel: in lossless conditions Liouville's Theorem ensures conservation of the
number of independent information channels during propagation,
while Gromov's Non-Squeezing
Theorem sets a geometric lower bound on resolution, asserting that
no Hamiltonian transformation can compress an information channel
below a minimum symplectic cell. While the Non-Squeezing
Theorem provides a powerful analogy for the limits of resolution,
its rigorous application to the inherently diffuse boundaries of
radiated fields remains an open mathematical challenge.

More broadly, the results point to a conclusion that goes beyond
NDF estimation: the spatial information content of an
electromagnetic field is not an intrinsic property of the field
itself, but is born from the geometric structure of the
source--observer configuration. In the asymptotic regime, the
physical complexity of the electromagnetic problem---the vector
nature of the field, the specific source distribution, the
near-field corrections---is progressively washed out, leaving only
a geometric skeleton governed by the symplectic structure of the
source--observer system. Symplectic geometry is the fabric of what
has been called the \textit{Deep Physical
	Layer}~\cite{migliore2021world}---the fundamental substrate on
which spatial information is woven into the electromagnetic field,
and whose topology determines what information structures are
possible and which are geometrically forbidden.

Recalling the metaphor of~\cite{migliore2018horse}, where
electromagnetics is the horse and information theory is the
horseman, the horse ultimately sets the limits regardless of the
rider. Yet even the most powerful horse is constrained by the
terrain it travels. Symplectic geometry is that terrain: the
horseman may choose his path, but the landscape---shaped by the
symplectic geometry of the source--observer configuration---%
determines which paths exist and which are forever closed.

	\section*{Acknowledgment}
	
	The author is indebted to Professor Ovidio Mario Bucci for his
	insightful discussions and valuable suggestions, which greatly enhanced
	the quality and rigor of this work, and to the three anonymous
	referees, whose comments and suggestions significantly improved the
	manuscript.

	\appendix
	
	\section{Equivalence of Coherent Modes and Singular Functions}
	\label{app:theorem}
	
	Throughout this appendix, $\dagger$ denotes the Hermitian adjoint;
	bold upright letters denote vector fields; double-bar symbols denote
	dyadics. All domains $D$ and $S_{\mathrm{obs}}$ are bounded,
	measurable, and disjoint subsets of $\mathbb{R}^3$.
	
	\smallskip
	\noindent\textbf{Theorem (Equivalence of Coherent Modes and Singular
		Functions).}\enspace
	\textit{%
		Let $\mathcal{A}: L^2(D;\mathbb{C}^3)\to
		L^2(S_{\mathrm{obs}};\mathbb{C}^3)$ be the radiation operator
		\begin{equation}
			(\mathcal{A}\bJ)(\br)
			= \int_D \dG(\br,\br')\,\bJ(\br')\,d\mu_D(\br'),
			\label{eq:app_A}
		\end{equation}
		where the dyadic Green function $\dG$ satisfies the
		Hilbert--Schmidt condition
		\begin{equation}
			\int_{S_{\mathrm{obs}}}\!\int_D
			\|\dG(\br,\br')\|_F^2\,
			d\mu_D(\br')\,d\mu_{S_{\mathrm{obs}}}(\br) < \infty.
			\label{eq:app_HS}
		\end{equation}
		Let $\{(\sigma_n,\bu_n)\}$ be the singular values and left singular
		functions of $\mathcal{A}$, and let $\bJ(\br')$ be a zero-mean,
		spatially incoherent, homogeneous stochastic source with
		cross-spectral density dyadic
		\begin{equation}
			\E\!\bigl[\bJ(\br'_1)\bJ(\br'_2)^\dagger\bigr]
			= \dI\,\delta(\br'_1-\br'_2).
			\label{eq:app_WJ}
		\end{equation}
		Then the CSD of the radiated field $\bE=\mathcal{A}\bJ$ satisfies
		\begin{equation}
			\E\!\bigl[\bE(\br_1)\bE(\br_2)^\dagger\bigr]
			= \int_D \dG(\br_1,\br')\,\dG(\br_2,\br')^\dagger\,d\mu_D(\br'),
			\label{eq:app_W}
		\end{equation}
		i.e., the CSD operator $\mathcal{W}=\mathcal{A}\mathcal{A}^\dagger$.
		Consequently, the coherent modes $\bpsi_n$ and eigenvalues $\lambda_n$
		of the Karhunen--Lo\`eve expansion satisfy
		$\bpsi_n = e^{i\phi_n}\bu_n$ and $\lambda_n = \sigma_n^2$
		for all $n\ge 1$.%
	}
	
	\smallskip
	\noindent\textbf{Proof.}\enspace
	Since $\mathcal{A}$ is Hilbert--Schmidt by~\eqref{eq:app_HS}, the CSD
	operator can be written through the source covariance $\Sigma_J=\dI$
	as $\mathcal{W}=\mathcal{A}\,\Sigma_J\,\mathcal{A}^\dagger$, which is
	trace class; this justifies exchanging expectation and integration.
	Applying this to
	$\E[(\mathcal{A}\bJ)(\br_1)(\mathcal{A}\bJ)(\br_2)^\dagger]$
	and using~\eqref{eq:app_WJ}:
	\begin{align}
		\E\!\bigl[\bE(\br_1)\bE(\br_2)^\dagger\bigr]
		&= \iint_{D^2}
		\dG(\br_1,\br'_1)\,\dI\,\delta(\br'_1-\br'_2)
		\,\dG(\br_2,\br'_2)^\dagger
		\,d\mu_D^2 \notag\\
		&= \int_D
		\dG(\br_1,\br')\,\dG(\br_2,\br')^\dagger
		\,d\mu_D(\br').
		\label{eq:app_proof}
	\end{align}
	The adjoint satisfies $(\mathcal{A}^\dagger\mathbf{f})(\br') =
	\int_{S_{\mathrm{obs}}}\dG(\br,\br')^\dagger\mathbf{f}(\br)
	\,d\mu_{S_{\mathrm{obs}}}(\br)$,
	so the kernel of $\mathcal{A}\mathcal{A}^\dagger$ equals the
	right-hand side of~\eqref{eq:app_proof}, giving
	$\mathcal{W}=\mathcal{A}\mathcal{A}^\dagger$ a.e.
	The spectral theorem for compact self-adjoint
	operators~\cite{kolmogorov1999elements} then yields the shared
	eigenbasis $\bpsi_n=e^{i\phi_n}\bu_n$ and $\lambda_n=\sigma_n^2$.
	\hfill$\blacksquare$
	
	\smallskip
	\noindent\textbf{Remark.}\enspace
	The result holds for any domain type ($dV'$, $dS'$, or $dl'$ for
	$d\mu_D$), provided~\eqref{eq:app_HS} holds---guaranteed when $D$ and
	$S_{\mathrm{obs}}$ are disjoint and bounded, since the free-space
	dyadic Green function is then analytic and square-integrable on
	$S_{\mathrm{obs}}\times D$. This establishes rigorously the
	equivalence stated in Section~\ref{sec:EIT_modelA}: the stochastic
	\NDF\ coincides with the deterministic \NDF\ for any spatially
	incoherent homogeneous source, in both the scalar and full vector
	settings.

	\bibliographystyle{IEEEtran}
	\bibliography{Biblio_SIFv3}
	
\end{document}